\documentclass[final]{siamltex}
\usepackage{epsfig}
\usepackage{graphics}
\usepackage{graphicx}

\title{Existence and Stability of Standing Pulses in Neural Networks:
  II. Stability} 

\author{Yixin Guo\thanks{Department of Mathematics, The Ohio State 
University, Columbus, OH 43210 ({\tt yigst@math.ohio-state.edu}).}
        \and Carson C. Chow\thanks{Department of Mathematics,
University of Pittsburgh, Pittsburgh, PA 15260 ({\tt ccchow@pitt.edu}).}}

\begin{document}

\maketitle

\begin{abstract}
We analyze the stability of standing pulse solutions of a neural
network integro-differential equation. The network consists of a
coarse-grained layer of 
neurons synaptically connected by lateral inhibition with a
non-saturating nonlinear gain function.  When two standing
single-pulse solutions coexist, the small pulse is unstable, and the
large pulse is stable.  The large single-pulse is bistable with the
``all-off'' state.  This bistable localized activity may have strong
implications for the mechanism underlying working memory.  We show
that dimple pulses have similar stability properties to large pulses but
double pulses are unstable.
\end{abstract}



\pagestyle{myheadings}
\thispagestyle{plain}
\markboth{Yixin Guo and Carson C. Chow}{Existence and Stability of
  Standing Pulses in Neural Networks: II. Stability}

\section{Introduction}
\label{sec:intro}

In the accompanying paper~\cite{Guo2}, we considered stationary
localized self-sustaining solutions of an integro-differential neural
network or neural field equation.  The pulses are bistable
with an inactive neural state and could be the underlying mechanism of
persistent neuronal activity responsible for working memory.  However,
in order to serve as a memory, these states must be stable to
perturbations.  Here we compute the linear stability of stationary
pulse states. 
 
The neural field equation has the form
\begin{equation}
\label{eq:int}
\frac{\partial{u(x,t)}}{\partial{t}}+u(x,t) =
\int_{-\infty}^{\infty}w(x-y)f[u(y)]dy 
\end{equation}
with a nonsaturating gain function
\begin{equation}
f[u]=[\alpha (u(y,t)-u_{\scriptscriptstyle
    T})+1]\Theta(u-u_{\scriptscriptstyle T}) 
\end{equation}
where $\Theta(\cdot)$ is the Heaviside function, 
and ``wizard hat'' connection function
\begin{equation}
\label{eq:coupling}
w(x)=Ae^{-a|x|}-e^{-|x|}.
\end{equation}

\begin{figure}[!htb]
\centering
\epsfig{figure=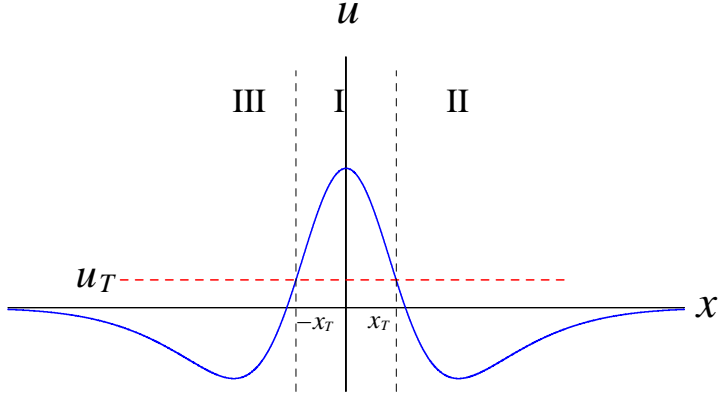,height=2in} 
\caption{\small Single-pulse solution.} 
\label{fig:bumppic}
\end{figure}

In Ref~\cite{Guo2}, we
considered stationary solutions $u_0(x)$, where
$u_0(x)>u_{\scriptscriptstyle T}$ on an interval 
$-x_{\scriptscriptstyle
T}<x<x_{\scriptscriptstyle
T}$, $u(x_{\scriptscriptstyle T},t)=u_{\scriptscriptstyle T}$, and
$u(x,t)=u_0(x)$ satisfies the stationary integral equation 
\begin{equation}
u_0(x)= \int_{-x_T}^{x_T}w(x-y)[\alpha
(u_0(y)-u_{\scriptscriptstyle T})+1]dy,
\label{eq:bump}
\end{equation}
We have shown the existence of pulse solutions of equation
(\ref{eq:bump}) in the  
form of single-pulses, dimple pulses and
double pulses~\cite{Guo1,Guo2}.   Examples can be seen in
Figs.~\ref{fig:bumppic}, \ref{fig:margindimple}  and
\ref{fig:amari2pulse}.
We constructed the pulses by converting the integral equation
(\ref{eq:bump}) into piecewise-linear ODEs and then matching their
solutions at the threshold points $x_{\scriptscriptstyle T}.$ When the
excitation $A$ and gain $\alpha$ is small,  there are no pulse
solutions.  If either is increased, there is a saddle-node bifurcation
where 
two coexisting single-pulses, a small one and a large one, arise.  As
the gain or excitation increases, 
more than two pulses can coexist.   For certain parameters, the large
pulse can  
become a dimple pulse, and  a dimple pulse can become a double
pulse~\cite{Guo1,Guo2}.   

In this paper, we derive an eigenvalue equation to analyze the stability
of the pulse solutions.  While our eigenvalue
equation is valid for any continuous and integrable connection
function $w(x)$, we explicitly compute the eigenvalues for
(\ref{eq:coupling}).  For the cases that we tested, we find that the
small pulse is unstable and the large pulse is stable.  If there is a
third (larger) pulse then it is unstable.  The stability properties of
dimple pulses are the same as corresponding large pulses.  Double
pulses are unstable. 

\section{Eigenvalue equation for stability} 
\label{sec:eigenvalueproblem}

We consider small perturbations around a stationary pulse solution by
substituting $u(x,t)=u_0(x)+\epsilon v(x,t)$ into (\ref{eq:int}),
where $\epsilon>0$ is 
small.  Since the pulse solutions are localized in
space, we must assume the perturbation to the pulse will lead to time
dependent changes to the boundaries of the stationary pulse (i.e where
$u_0(x_{\scriptscriptstyle T} )=u_{\scriptscriptstyle T}$).
Thus the boundaries $-x_{\scriptscriptstyle T}$ and
$x_{\scriptscriptstyle T}$ become time dependent functions
\begin{eqnarray}
x_1(t)&=&-x_{\scriptscriptstyle T}+\epsilon\Delta_1(t)\label{eq:x1}\\
x_2(t)&=&x_{\scriptscriptstyle T}+\epsilon\Delta_2(t)\label{eq:x2}
\end{eqnarray}
where $\epsilon\Delta_1(t)$ and $\epsilon\Delta_2(t)$ are the changes of the 
boundaries $-x_{\scriptscriptstyle T}$ and $x_{\scriptscriptstyle T}$ produced
by the small perturbations.
Inserting $u(x,t)=u_0(x)+\epsilon v(x,t)$ into (\ref{eq:int}) and eliminating
the stationary solution with (\ref{eq:bump}) gives
\begin{eqnarray}
\label{eq:eigenvalueequation2} v_t(x,t)+v(x,t)=\alpha
\int_{-x_{\scriptscriptstyle T}}^{x_{\scriptscriptstyle
T}}w(x-y)v(y,t)dy + I_1+I_2
\end{eqnarray}
where
\begin{eqnarray}
I_1 & = & \int_{-(x_{\scriptscriptstyle
    T}+\epsilon\Delta_1(t))}^{-x_{\scriptscriptstyle T}}w(x-y)[\alpha 
(u_0(y)+\epsilon v(y,t)-u_{\scriptscriptstyle T})+1]dy, \\
I_2 & = & \int_{x_{\scriptscriptstyle T}}^{x_{\scriptscriptstyle
    T}+\epsilon\Delta_2(t)}w(x-y)[\alpha 
(u_0(y)+\epsilon v(y,t)-u_{\scriptscriptstyle T})+1]dy.
\end{eqnarray}
Expanding the integrals $I_1$ and $I_2$ to order $\epsilon$ yields the
linearized dynamics for the perturbations $v(x,t)$
\begin{equation}
\label{eq:eigenvalueequation3} 
\hspace{0.8cm} v_t(x,t)+v(x,t)=\alpha
\int_{-x_{\scriptscriptstyle T}}^{x_{\scriptscriptstyle
T}}w(x-y)v(y,t)dy -w(x+x_{\scriptscriptstyle T}) \Delta_1
+w(x-x_{\scriptscriptstyle T}) \Delta_2.
\end{equation}

The time dependence of $\Delta_1$ and $\Delta_2$ is found by using the
fact that $u(x,t)$ is equal to the threshold $u_{\scriptscriptstyle
  T}$ at the boundaries 
of the pulse.  Inserting (\ref{eq:x1}) and (\ref{eq:x2}) into the
boundary condition $u(x_1(t),t)=u_{\scriptscriptstyle T}$ and
expanding to first order in 
$\epsilon$ leads to
\begin{equation}
\Delta_1(t)=-\frac{v(-x_{\scriptscriptstyle
    T},t)}{c}
\label{delta1}
\end{equation}
where 
\begin{equation}
  c=\left.\frac{du_0(x)}{dx}\right|_{x=-x_{\scriptscriptstyle T}}>0.
\end{equation}
Similarly, 
\begin{equation}
\Delta_2(t)=\frac{v(x_{\scriptscriptstyle T},t)}{c}.
\label{delta2}
\end{equation}
Consider time variations of $v(x,t)$ that obey 
\begin{equation}
v(x,t)= v(x)e^{\lambda t}
\label{eq:v}
\end{equation}
 where 
$v(x)$ is a bounded and
continuous function that decays to $0$ exponentially as $x \rightarrow
\pm \infty$.  Substitute (\ref{eq:v}) with (\ref{delta1}) and (\ref{delta2})
into
(\ref{eq:eigenvalueequation3}), to obtain 
\begin{equation}
\label{eq:eigenvalueequation} 
\hspace{1cm}(1+\lambda)v(x)= w(x-x_{\scriptscriptstyle
T})\frac{v(x_{\scriptscriptstyle T})}{c}+w(x+x_{\scriptscriptstyle
T})\frac{v(-x_{\scriptscriptstyle
T})}{c}+\alpha\int_{-x_{\scriptscriptstyle T}}^{x_{\scriptscriptstyle
T}}w(x-y)v(y)dy.
\end{equation}  
where $\lambda$ is an eigenvalue with corresponding eigenfunction
$v(x)$.  Equation (\ref{eq:eigenvalueequation}) is an eigenvalue
problem that governs the stability of small perturbations to pulse
solutions of the neural field equation (\ref{eq:int}).  If the real
parts of all the eigenvalues are negative, the stationary pulse
solution $u_0(x)$ is stable. If the real part of one of the
eigenvalues is positive, $u_0(x)$ is unstable.

We define an operator $L$: $C\left[-x_{\scriptscriptstyle
T},x_{\scriptscriptstyle T}\right] \rightarrow
C\left[-x_{\scriptscriptstyle T},x_{\scriptscriptstyle T}\right]$:
\begin{eqnarray}
\label{eq:operatorL}
L v(x)=w(x-x_{\scriptscriptstyle T})\frac{v(x_{\scriptscriptstyle
T})}{c}+w(x+x_{\scriptscriptstyle T})\frac{v(-x_{\scriptscriptstyle
T})}{c}+\alpha \int_{-x_{\scriptscriptstyle T}}^{x_{\scriptscriptstyle
T}}w(x-y)v(y)dy.
\end{eqnarray}
Then the eigenvalue equation (\ref{eq:eigenvalueequation}) becomes
\begin{eqnarray}
\label{eq:operatoreqn}
(1+\lambda)v(x)=L(v(x)) \mbox{\hspace{1.5cm} on\hspace{0.5cm}}
C\left[-x_{\scriptscriptstyle T},x_{\scriptscriptstyle T}\right].
\end{eqnarray}

We show in the Appendix (Theorem \ref{the:theapp}) that  $L$ is a
compact operator.  We also show the following 
properties of the eigenvalue equation (\ref{eq:eigenvalueequation}): 
\vspace{.25cm}
\begin{enumerate}
\item Eigenvalues $\lambda$ are always real (Theorem~\ref{the:the1}).
\item Eigenvalues $\lambda$ are bounded by ${\displaystyle \lambda_b
\equiv \frac{2k_0}{c}+ 2 \alpha k_1 x_{\scriptscriptstyle T}-1}$ where
$k_0$ is the maximum of $\left|w(x)\right|$ on $\left[0,
2x_{\scriptscriptstyle T}\right]$ and $\left|w(x-y)\right|\leq k_1$
for all $(x,y) \in J \times J,$ $J=\left[-x_{\scriptscriptstyle
T},x_{\scriptscriptstyle T}\right]$ (Theorem \ref{the:the2}).
\item Zero is always an eigenvalue (Theorem \ref{the:the9}).
\item $\lambda=-1$ is the only possible accumulation point of the
eigenvalues (Theorem \ref{the:the5}).  Thus, the only possible
essential spectrum of 
operator $L$ is located at $\lambda=-1$ implying that the discrete
spectrum of $L$ (i.e. eigenvalues of
(\ref{eq:eigenvalueequation})) captures all of the stability
properties.
\end{enumerate}
We use these properties to compute the discrete eigenvalues to determine
stability of the pulse solutions.

\section{Linear stability analysis of the Amari case ($\alpha=0$)}
\label{sec:amarilinearstability}

Amari~\cite{amari77} computed the stability of pulse solutions to
(\ref{eq:int}) for $\alpha=0$.  He obtained stability by computing the
dynamics of the pulse boundary points.  He found that the small pulse
is always unstable and the large pulse is always stable.  Pinto and
Ermentrout~\cite{Pinto1} later confirmed Amari's results by deriving
an eigenvalue problem for small perturbations.

We consider a stationary pulse solution of (\ref{eq:int}) with width
$x_{\scriptscriptstyle T}$.  Applying eigenvalue equation
(\ref{eq:eigenvalueequation}) to 
the Amari case yields 
\begin{eqnarray}
(1+\lambda)v(x)=w(x-x_{\scriptscriptstyle
T})\frac{v(x_{\scriptscriptstyle T})}{c}+w(x+x_{\scriptscriptstyle
T})\frac{v(-x_{\scriptscriptstyle T})}{c}\equiv T_1(v(x)),
\label{amarilin}
\end{eqnarray}
where $T_1$ is a compact operator on $C[-x_{\scriptscriptstyle T}, 
x_{\scriptscriptstyle T}]$ (see Theorem \ref{the:theapp}).  The
spectrum of a compact operator is a countable set with no accumulation
point different from zero. Therefore, the only possible location of
the essential spectrum for $T_1$ is at $\lambda=-1$.  This implies
that instability of a pulse is indicated by the existence of a
positive discrete eigenvalue.

The eigenvalue $\lambda$ can be obtained by
setting  $x=-x_{\scriptscriptstyle T}$ and
$x=x_{\scriptscriptstyle T}$ in (\ref{amarilin}) to give a two
dimensional system
\begin{eqnarray}
\left(1+\lambda-\frac{w(0)}{c}\right)v(x_{\scriptscriptstyle
T})-\frac{w(2x_{\scriptscriptstyle T})}{c}v(-x_{\scriptscriptstyle
T})=0\label{2by2a}\\
-\frac{w(2x_{\scriptscriptstyle T})}{c}v(x_{\scriptscriptstyle
T})+\left(1+\lambda-\frac{w(0)}{c}\right)v(-x_{\scriptscriptstyle
  T})=0
\label{2by2b}
\end{eqnarray}
This is identical to the eigenvalue equation of Ref \cite{Pinto1}.
Setting the determinant of system
(\ref{2by2a}) and (\ref{2by2b}) to zero
gives the eigenvalues
\begin{equation}
\lambda=\frac{w(0) \pm w(2x_{\scriptscriptstyle T})}{c}-1,
\label{eiv}
\end{equation} 
which agrees with Ref.~\cite{Pinto1}. 

The stationary solution of the Amari problem satisfies
\begin{equation}
u(x)=\int_{-x_{\scriptscriptstyle T}}^{x_{\scriptscriptstyle T}}w(x-y)dy=
\int_{x+x_{\scriptscriptstyle T}}^{x-x_{\scriptscriptstyle T}}w(y)dy
\end{equation}
Differentiating $u(x)$ yields
$u'(x)=w(x+x_{\scriptscriptstyle T})-w(x-x_{\scriptscriptstyle T})$ implying 
\begin{equation}
u'(-x_{\scriptscriptstyle
T})=w(0)-w(2x_{\scriptscriptstyle T})=c.
\end{equation}
Inserting into (\ref{eiv}) gives the eigenvalues
\begin{equation}
\lambda=\frac{w(0)+w(2x_{\scriptscriptstyle T})}{c}-1, \ \ 0
\label{eiv2}
\end{equation}
The zero eigenvalue was expected from translational symmetry.  Since
$w(0)>w(2x_T)$, the sign of $c$
alone determines stability of the pulse.  Recall that the small and
large pulse arise from a saddle node
bifurcation~\cite{amari77,Guo1,Guo2}.  At the saddle node bifurcation,
both eigenvalues are zero.  Thus, setting $\lambda=0$ in (\ref{eiv2})
shows that 
the width of the pulse satisfies $w(2x_{\scriptscriptstyle
  T})=0$~\cite{amari77}. 
For our
connection function, $w(x)$ has only one zero at $x_0$ for $w(x)$ on
$(0, \infty)$ (see \cite{Guo1, Guo2}).  Thus $ x_{\scriptscriptstyle
T}=x_0/2$ at the saddle node.  For the large pulse, $
x_{\scriptscriptstyle T}>x_0/2$, implying $w(2x_{\scriptscriptstyle
T})<0$ and $c>0$.  Conversely, $c<0$ for the small pulse.  Thus the
large pulse is stable and small pulse is unstable.

Consider the example: $a=2.4$, $A=2.8$, $u_{\scriptscriptstyle T}=0.400273$, 
$\alpha=0.$ There exist two single-pulses, the large pulse {\bf l} and 
the small pulse {\bf s}~\cite{Guo1, Guo2}. For
the pulse {\bf l},  
 $x_{\scriptscriptstyle T}^{\bf l}=0.607255$ gives the non-zero eigenvalue 
$\lambda=-0.165986<0$, indicating it is stable. For the small pulse {\bf s},
$x_{\scriptscriptstyle T}^{\bf s}=0.21325,$
gives $\lambda=0.488339>0$ indicating it is
unstable.   

\section{Computing the eigenvalues}

For the case of $\alpha>0$, we must compute the eigenvalues of
(\ref{eq:eigenvalueequation}) with the integral
operator. 
Our strategy is to reduce the integral equation to a piecewise
linear ODE on three separate regions.  The discrete spectrum can then
be obtained from the zeros of the determinant of a linear system based
on the matching conditions between the regions. 
This approach is similar to the Evans function
method~\cite{Evans,E1,E2,E3}. 

\subsection{ODE form of the eigenvalue problem}
\label{sec:odereduce} We transform
(\ref{eq:eigenvalueequation}) (with the connection function defined by
(\ref{eq:coupling})) into three piecewise linear
ordinary differential equations (ODEs) on $(-\infty,
x_{\scriptscriptstyle T_1})$, $(-x_{\scriptscriptstyle T_1},
x_{\scriptscriptstyle T_1})$ and $(-x_{\scriptscriptstyle T_1},\infty)$.  
The ODEs then obey a set of
matching conditions at $x=x_{\scriptscriptstyle T_1}$ and
$x=-x_{\scriptscriptstyle T_1}$.  

On the domain  $x \in (-x_{\scriptscriptstyle T_1},
x_{\scriptscriptstyle T_1})$, we can write (\ref{eq:eigenvalueequation})
in the form  
\begin{eqnarray}
\label{eq:localeq9} (1+\lambda)v(x)=T_1(x)+I_1-I_2+I_3-I_4
\end{eqnarray}
where
\begin{eqnarray*}
I_1(x) =  \alpha \int_{-x_{\scriptscriptstyle
    T_1}}^{x}Ae^{-a(x-y)}v(y)dy, & \mbox{\hspace{1cm}}& 
I_2(x)  =  \alpha \int_{-x_{\scriptscriptstyle T_1}}^{x}e^{-(x-y)}v(y)dy \\
I_3(x)  =  \alpha \int_{x}^{x_{\scriptscriptstyle
T_1}}Ae^{a(x-y)}v(y)dy, & \mbox{\hspace{1cm}} & I_4(x)  = \alpha
\int_{x}^{x_{\scriptscriptstyle T_1}}e^{(x-y)}v(y)dy
\end{eqnarray*}
and
\begin{equation}
\label{eq:teqn}
T_1(x) ={\displaystyle  w(x-x_{\scriptscriptstyle
T_1})\frac{v(x_{\scriptscriptstyle T_1})}{c}+w(x+x_{\scriptscriptstyle
T_1})\frac{v(-x_{\scriptscriptstyle T_1})}{c}}.
\end{equation}

Differentiating (\ref{eq:localeq9}) repeatedly gives
\begin{eqnarray}
\hspace{-1.2cm} \label{eq:localeq10}
(1+\lambda)v'(x) &=&T_1'(x)-aI_1+I_2+aI_3-I_4 \\
(1+\lambda)v''(x)&=&T_1''(x)+a^2I_1-I_2+a^2I_3-I_4+2 \alpha (1-aA)v(x) \\
\label{eq:localeq11}
(1+\lambda)v'''(x)&=&T_1'''(x)-a^3I_1+I_2+a^3I_3-I_4+2 \alpha
(1-aA)v'(x) \\
\label{eq:localeq12}
\hspace{-1cm}(1+\lambda)v^{''''}(x)&=&T_1^{''''}(x)+a^4I_1-I_2+a^4I_3-I_4+2
\alpha (1-a^3A)v(x)+ \\ 
& & \mbox{}2 \alpha (1-aA)v''(x) \nonumber
\end{eqnarray}
where we have used
\begin{eqnarray*}
I_1' = -a I_1 + \alpha A v(x), & \mbox{\hspace{1.5cm}}&
I_2' = -I_2+\alpha v(x) \\
I_3' = a I_3 - \alpha A v(x), & \mbox{\hspace{1.5cm}}& I_4' = I_4-\alpha v(x).
\end{eqnarray*}
Taking $(\ref{eq:localeq11})-a^2(\ref{eq:localeq9})$ and rearranging gives
\begin{equation}
\label{eq:I2I4}
I_2+I_4 = \frac{1}{a^2-1}\left[(\lambda+1)v''+(2 \alpha a A-2
\alpha-a^2 \lambda -a^2)v+a^2T_1-T_1'' \right].
\end{equation}
Substituting (\ref{eq:I2I4}) back into (\ref{eq:localeq9}) leads to
\begin{equation}
\label{eq:I1I3}
I_1+I_3 =  \frac{1}{a^2-1}\left[(\lambda+1)v''+(2 \alpha a A-2\alpha
-\lambda-1)v+T_1-T_1'' \right].
\end{equation}
Substituting both (\ref{eq:I2I4}) and (\ref{eq:I1I3}) into
(\ref{eq:localeq12}), results in a fourth order ordinary
differential equation for $v$ on the domain $x \in (-x_{\scriptscriptstyle
T},x_{\scriptscriptstyle T})$
\begin{eqnarray}
\label{eq:stabilityode1}
\lefteqn{\hspace{-8cm} \frac{1+\lambda}{\alpha}v^{''''}
 =\left[\frac{(1+\lambda)(a^2+1)}{\alpha}+2(1-aA)\right]v''
 +a\left[2(A-a)-\frac{\lambda+1}{\alpha}a \right]v +}\\ 
 T_1^{''''}(x)-(1+a^2)T_1''(x)+a^2 T_1(x). \nonumber 
\end{eqnarray}
Using $T_1^{''''}(x)-(1+a^2)T_1''(x)+a^2 T_1(x)=0$ (obtained by differentiating
$T_1(x)$) and simplifying, leads to 
\begin{equation}
 \label{eq:stabilityode1b}
(1+\lambda)v^{''''} -Bv''+Cv=0 , \qquad x\in (-x_{\scriptscriptstyle
   T}, x_{\scriptscriptstyle T})
\end{equation}
 where $B=(1+\lambda)(a^2+1)+2
\alpha(1-aA),$ and $C=(\lambda+1)a^2-2 \alpha a(A-a)$.

On the domain  $x\in (x_{\scriptscriptstyle T}, \infty)$,
(\ref{eq:eigenvalueequation}) 
can be written as
\begin{equation}
\label{eq:localeq17} 
(1+\lambda)v=T_1+ J_1-J_2,
\end{equation}
where
$$J_1=\alpha A\int_{-x_{\scriptscriptstyle T}}^{x_{\scriptscriptstyle
    T}}e^{-a(x-y)}v(y)dy, \mbox{\hspace{1cm}} 
J_2=\int_{-x_{\scriptscriptstyle T}}^{x_{\scriptscriptstyle
    T}}e^{-(x-y)}v(y)dy,$$ 
and $T_1$ is defined by (\ref{eq:teqn}) on the  domain $(x_{\scriptscriptstyle
T}, \infty)$.

Differentiating (\ref{eq:localeq17})  and using $J_1'=-a J_1$
and $J_2'=-J_2$ gives
\begin{eqnarray}
(1+\lambda)v'(x)& = & T_1'-a J_1+J_2, \label{eq:vprime}\\
(1+\lambda)v''(x)& = & T_1''+a^2 J_1-J_2. \label{eq:vprime2}
\end{eqnarray}
Taking
$a$(\ref{eq:localeq17})+$(a+1)$(\ref{eq:vprime})+(\ref{eq:vprime2})
and using $T_1''+(1+a)T_1'+ aT_1=0$ leads to 
\begin{equation}
v''+(a+1)v'+a v =0, \qquad x\in (x_{\scriptscriptstyle T}, \infty).
\end{equation}
Similarly, the ODE on $(-\infty, -x_{\scriptscriptstyle T})$ is given by
\begin{equation}
\label{eq:ode1}
v''-(a+1)v'+a v =0, \qquad x\in (-\infty, -x_{\scriptscriptstyle T}).
\end{equation}
In summary, the eigenvalue problem (\ref{eq:eigenvalueequation})
reduces to three ODEs: 
\begin{eqnarray*} 
&& ({\rm ODE~I}) \ \ \qquad \qquad v''-(a+1)v'+a v =0, \qquad \qquad
  x\in (-\infty, 
  -x_{\scriptscriptstyle T}),\\ 
&& ({\rm ODE~II}) \ \qquad \qquad (1+\lambda)v^{''''}-B v''+C v=0, \
  \qquad    x\in 
  (-x_{\scriptscriptstyle T}, x_{\scriptscriptstyle  T}), \\ 
&& ({\rm ODE~III}) \qquad  \qquad v''+(a+1)v'+av=0, \qquad \qquad x\in
  (x_{\scriptscriptstyle T}, \infty), 
\end{eqnarray*}
where $B=(1+\lambda)(a^2+1)+2\alpha(1-aA)$ and
$C=(\lambda+1)a^2-2\alpha a(A-a)$.

\subsection{Matching Conditions}
\label{sec:discontinuity}

The solutions of ODE I, II, and III and their first three derivatives
must satisfy a set of matching conditions across the boundary points
$-x_{\scriptscriptstyle T}$ and $x_{\scriptscriptstyle T}$.  We derive
these conditions from the original eigenvalue equation
(\ref{eq:eigenvalueequation}) 
which we write as
\begin{eqnarray}
\label{eq:localeq8} c(1+\lambda)v(x)= w(x-x_{\scriptscriptstyle
T})v(x_{\scriptscriptstyle T})+w(x+x_{\scriptscriptstyle
T})v(-x_{\scriptscriptstyle T})+c \alpha W(x),
\end{eqnarray}
where
$W(x)={\displaystyle \int_{-x_{\scriptscriptstyle
T}}^{x_{\scriptscriptstyle T}}w(x-y)v(y)dy},$ $x \in (-\infty,
\infty)$.
{}From (\ref{eq:localeq8}), we see that $v(x)$ is continuous on
$(-\infty, \infty)$.  
However,  $w(x)$ has a cusp at $ x=0$ which will lead to
discontinuities in the derivatives of $v(x)$ across the boundary
points 
$-x_{\scriptscriptstyle T}$ and $x_{\scriptscriptstyle T}$. 

We first probe the discontinuities of $W(x)$ and its derivatives. 
$W(x)$ is continuous on $(-\infty, \infty).$
By a change of variables, 
$W(x)=\int_{x-x_{\scriptscriptstyle
T}}^{x+x_{\scriptscriptstyle T}}w(z)v(x-z)dz$,
from which we obtain
\begin{eqnarray*}
W'(x)= w(x+x_{\scriptscriptstyle T})v(-x_{\scriptscriptstyle
T})-w(x-x_{\scriptscriptstyle T})v(x_{\scriptscriptstyle T})
+\int_{x-x_{\scriptscriptstyle
T}}^{x+x_{\scriptscriptstyle T}}w(z)v'(x-z)dz,
\end{eqnarray*}
indicating that $W'(x)$ is also continuous on $(-\infty, \infty)$.
However $W'(x)$ is not smooth at $-x_{\scriptscriptstyle T}$ and
$x_{\scriptscriptstyle T}$. 
Differentiating $W'(x)$ for $x \neq -x_{\scriptscriptstyle
T}, x_{\scriptscriptstyle T}$ gives
\begin{eqnarray*}
W''(x) & = & w'(x+x_{\scriptscriptstyle
T})v(-x_{\scriptscriptstyle T})-w'(x-x_{\scriptscriptstyle
T})v(x_{\scriptscriptstyle T})+w(x+x_{\scriptscriptstyle
T})v'(-x_{\scriptscriptstyle T}^{+})\\
& - & w(x-x_{\scriptscriptstyle T})v'(x_{\scriptscriptstyle T}^{-}) 
- \int_{x+x_{\scriptscriptstyle T}}^{x-x_{\scriptscriptstyle
T}}w(z)v''(x-z)dz
\end{eqnarray*}
where ${\displaystyle v'(-x_{\scriptscriptstyle
    T}^{+})=\lim_{x\rightarrow -x_T^+} v'(x)}$ for
$x>-x_{\scriptscriptstyle T}$ (right limit) and 
${\displaystyle v'(x_{\scriptscriptstyle T}^{-})=\lim_{x\rightarrow
    x_T^-} v'(x)}$ for $x<x_{\scriptscriptstyle T}$ (left limit).  

Using the following convention:
$$\left[ \cdot \right]|_{x=x_{\scriptscriptstyle
    T}}=\cdot|_{x=x_{\scriptscriptstyle T}^{+}}-
\cdot|_{x=x_{\scriptscriptstyle T}^{-}} \mbox{\hspace{1cm}} 
\left[ \cdot \right]|_{x=-x_{\scriptscriptstyle
    T}}=\cdot|_{x=-x_{\scriptscriptstyle T}^{+}}-
    \cdot|_{x=-x_{\scriptscriptstyle T}^{-}} $$ 
to represent the jump at the
boundaries,
we find that
\begin{eqnarray*}
\left[W''(x_{\scriptscriptstyle
T})\right] & = & W''(x)|_{x=x_{\scriptscriptstyle T}^{+}}-
W''(x)|_{x=x_{\scriptscriptstyle T}^{-}} =
-\left[w'(0)\right]v(x_{\scriptscriptstyle 
T})\\ 
\left[W''(-x_{\scriptscriptstyle
T})\right] & = &  W''(x)|_{x=-x_{\scriptscriptstyle T}^{+}}-
W''(x)|_{x=-x_{\scriptscriptstyle T}^{-}} =
\left[w'(0)\right]v(-x_{\scriptscriptstyle T}) 
\end{eqnarray*}
We differentiate $W''(x)$ for $x \neq
-x_{\scriptscriptstyle T}, x_{\scriptscriptstyle T}$ and find
\begin{eqnarray*}
\left[W'''(x_{\scriptscriptstyle
T})\right] & = & -\left[w''(0)\right]v(x_{\scriptscriptstyle
T})-\left[w'(0)\right]v'(x_{\scriptscriptstyle T}^-)\\
\left[W'''(-x_{\scriptscriptstyle
T})\right] & = & \left[w''(0)\right]v(-x_{\scriptscriptstyle
T})+\left[w'(0)\right]v'(-x_{\scriptscriptstyle T}^+) 
\end{eqnarray*}

To find the matching conditions for the derivatives of $v(x)$, we
differentiate (\ref{eq:localeq8}) with respect to $x$ for $x \neq
-x_{\scriptscriptstyle T},  x_{\scriptscriptstyle T}$,
and obtain
\begin{eqnarray*}
c(1+\lambda)v'(x)= w'(x-x_{\scriptscriptstyle
T})v(x_{\scriptscriptstyle T})+w'(x+x_{\scriptscriptstyle
T})v(-x_{\scriptscriptstyle T})+c \alpha W'(x).
\end{eqnarray*}
$v'(x)$ is discontinuous at the boundaries because of the
discontinuity of $w'(x)$ at $x=0$. Therefore
\begin{eqnarray*}
\left[v'(x_{\scriptscriptstyle T})\right] & = &
\frac{1}{c(1+\lambda)}\left[w'(0)\right]v(x_{\scriptscriptstyle
T}), \\
\left[v'(-x_{\scriptscriptstyle T})\right] & = & \frac{1}{c(1+\lambda)}\left[w'(0)\right]v(-x_{\scriptscriptstyle T}).
\end{eqnarray*}
Differentiating (\ref{eq:localeq8}) twice yields
\begin{eqnarray*}
 c(1+\lambda)v''(x)& = & w''(x-x_{\scriptscriptstyle
T})v(x_{\scriptscriptstyle T})+w''(x+x_{\scriptscriptstyle
T})v(-x_{\scriptscriptstyle T})+ c \alpha W''(x)
\mbox{\hspace{0.6cm}} x \neq -x_{\scriptscriptstyle
T},x_{\scriptscriptstyle T}
\end{eqnarray*}
\noindent There are discontinuities of $v''(x)$ at $-x_{\scriptscriptstyle T}$ and
$x_{\scriptscriptstyle T}$ that come from $W''(-x_{\scriptscriptstyle T})$ and $W''(x_{\scriptscriptstyle T})$. Note that $w''(0^-)=w''(0^+)$. 
The jump conditions of $v''(x)$ at $-x_{\scriptscriptstyle T}$ and
$x_{\scriptscriptstyle T}$ are
\begin{eqnarray*}
\left[v''(x_{\scriptscriptstyle T})\right] & = & \frac{\alpha}{1+\lambda}\left[W''(x_{\scriptscriptstyle T})\right] =-
\frac{\alpha}{1+\lambda}\left[w'(0)\right]v(x_{\scriptscriptstyle
T}),\\
\left[v''(-x_{\scriptscriptstyle T})\right] & = & \frac{\alpha}{1+\lambda}\left[W''(-x_{\scriptscriptstyle T})\right]= \frac{\alpha}{1+\lambda}\left[w'(0)\right]v(-x_{\scriptscriptstyle T}).
\end{eqnarray*}
By differentiating a third time we find the jump conditions for $v'''(x)$ at $-x_{\scriptscriptstyle T}$ and
$x_{\scriptscriptstyle T}$:
\begin{eqnarray*}
\left[v'''(x_{\scriptscriptstyle T})\right] & = &
 \frac{1}{c(1+\lambda)}\left[w'''(0)\right]v(x_{\scriptscriptstyle
T})+\frac{\alpha}{1+\lambda}\left[W'''(x_{\scriptscriptstyle T})\right] \\
& = & \frac{1}{c(1+\lambda)}\left[w'''(0)\right]v(x_{\scriptscriptstyle
T})-\frac{\alpha}{1+\lambda}\left[w'(0)\right]v'(x_{\scriptscriptstyle
T}^-), \\
\left[v'''(-x_{\scriptscriptstyle T})\right] & = & 
\frac{1}{c(1+\lambda)}\left[w'''(0)\right]v(-x_{\scriptscriptstyle
T})
+\frac{\alpha}{1+\lambda}\left[W'''(x_{\scriptscriptstyle T})\right] \\
& = & \frac{1}{c(1+\lambda)}\left[w'''(0)\right]v(-x_{\scriptscriptstyle
T})
+\frac{\alpha}{1+\lambda}\left[w'(0)\right]v'(-x_{\scriptscriptstyle T}^+)
\end{eqnarray*}
Using the connection function $w(x)$ defined in (\ref{eq:coupling}), we have 
\begin{eqnarray*}
\left[w'(0)\right]& = & w'(0^+)-w'(0^-)=2(1-a A) \\
\left[w''(0)\right]& = & w''(0^+)-w''(0^-)=0 \\
\left[w'''(0)\right]& = & w'''(0^+)-w'''(0^-)=2(1-a^3A) 
\end{eqnarray*}
These results lead directly to the following theorem.
\begin{theorem}
\label{the:the3} The continuous eigenfunction $v(x)$ on $(-\infty,
\infty)$ in (\ref{eq:eigenvalueequation}) has the following jumps
in its first, second and third order derivatives at
the boundary $-x_{\scriptscriptstyle T}$ and
$x_{\scriptscriptstyle T}$.
\begin{eqnarray}
\label{eq:R1}
\left[v( x_{\scriptscriptstyle T})\right] & = & 0\\
\label{eq:R2}
\left[v'(x_{\scriptscriptstyle T})\right] & = &
\frac{2 \alpha (1-a A)}{1+\lambda}v(x_{\scriptscriptstyle T})\\
\label{eq:R3}
\left[v''( x_{\scriptscriptstyle T})\right] & = &
\frac{2(a A-1)}{c(1+\lambda)}v(x_{\scriptscriptstyle T})\\
\label{eq:R4}
\left[v'''( x_{\scriptscriptstyle T})\right] & = &
\frac{2(1-a^3A)}{c(1+\lambda)}v(x_{\scriptscriptstyle T}) +
\frac{2 \alpha (aA-1)}{1+\lambda}v'(x_{\scriptscriptstyle T}^{-})
\end{eqnarray}
\begin{eqnarray}
\label{eq:L1}
\left[v(- x_{\scriptscriptstyle T})\right] & = & 0\\
\label{eq:L2}
\left[v'(- x_{\scriptscriptstyle T})\right] & = &
\frac{2 \alpha (1-a A)}{1+\lambda}v(- x_{\scriptscriptstyle T})\\
\label{eq:L3}
\left[v''(-x_{\scriptscriptstyle T})\right] & = &
\frac{-2(a A-1)}{c(1+\lambda)}v(- x_{\scriptscriptstyle T})\\
\label{eq:L4}
\left[v'''(- x_{\scriptscriptstyle T})\right] & = &
\frac{2(1-a^3A)}{c(1+\lambda)}v(-x_{\scriptscriptstyle T}) -
\frac{2 \alpha (aA-1)}{1+\lambda}v'(- x_{\scriptscriptstyle T}^{+}).
\end{eqnarray}
\end{theorem}

\subsection{Eigenfunction symmetries}
\label{sec:properties}

\begin{figure}[t]
\centering
\epsfig{figure=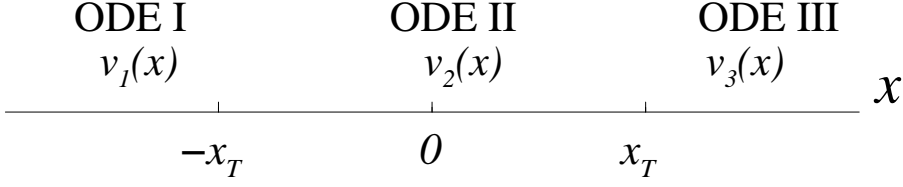,height=2.5cm} 
\caption{Valid ODEs on different sections and their solutions}
\label{fig:odesection0}
\end{figure}

We define  $v_1(x)$, $v_2(x)$ and $v_3(x)$ as the solutions of
ODE I, ODE II and ODE III, respectively (see Fig. \ref{fig:odesection0}.)
The three ODEs are all linear with constant 
coefficients. The continuous and bounded eigenfunction $v(x)$ of
(\ref{eq:eigenvalueequation}) is defined as the 
following
\begin{eqnarray*}v(x)=\left \{\begin{array}{ll} v_1(x), & \qquad 
x\in (-\infty, -x_{\scriptscriptstyle T}], \\
v_2(x), & \qquad x\in [-x_{\scriptscriptstyle T}, x_{\scriptscriptstyle T}], \\
v_3(x), & \qquad x \in [x_{\scriptscriptstyle T},\infty),
\end{array}\right.\end{eqnarray*}
and
$v_1(x)$ matches $v_2(x)$ at $-x_{\scriptscriptstyle T}$ and
$v_2(x)$ matches $v_3(x)$ at $x_{\scriptscriptstyle T}.$

\begin{lemma}
\label{pro:the6} The eigenfunction $v(x)$ is either even or odd.
\end{lemma}
\begin{proof}
By symmetry of ODE II, if $v_2(x)$ is a solution then 
$v_2(-x)$ is also a solution  Hence, both the even function ${\displaystyle
\frac{v_2(x)+v_2(-x)}{2}}$ and the odd function ${\displaystyle
\frac{v_2(x)-v_2(-x)}{2}}$ are solutions of ODE II.

Let
$$T_2(x)=\alpha \int_{-x_{\scriptscriptstyle
T}}^{x_{\scriptscriptstyle T}}w(x-y)v_2(y)dy.$$ 
If $v_2(x)$ is an even function, then since $w(x)$ is even, $T_2(x)$
is also even. 

By the continuity of $v(x)$ on $\Re$, $v(x_{\scriptscriptstyle T})$ and 
$v(-x_{\scriptscriptstyle T})$ can be replaced by
$v_2(x_{\scriptscriptstyle T}^-)$  
and $v_2(-x_{\scriptscriptstyle T}^+),$ respectively. 
Thus the eigenvalue problem (\ref{eq:eigenvalueequation}) is
\begin{eqnarray}
\label{eq:eigenvalueequation1}
(1+\lambda) v(x)= w(x-x_{\scriptscriptstyle T})\frac{v_2(x_{\scriptscriptstyle
T})}{c}+w(x+x_{\scriptscriptstyle
T})\frac{v_2(-x_{\scriptscriptstyle T})}{c}+T_2(x)
\end{eqnarray}
Given that
$v_2(x),$ $w(x)$ and $T_2(x)$ are all even functions, from
(\ref{eq:eigenvalueequation1}), we see that $v(x)$ is also even.
Similar, we can show that $v(x)$ is odd when $v_2(x)$ is odd. 
\end{proof}
\begin{lemma}
\label{lem:lem7} The matching conditions at $-x_{\scriptscriptstyle
  T}$ are identical to those at $x_{\scriptscriptstyle T}$ when $v(x)$
is an odd or an even function. 
\end{lemma}
\begin{proof}
This is shown with a  direct calculation of the matching conditions of
$v'(x)$, $v''(x)$ and $v'''(x)$ at both 
$-x_{\scriptscriptstyle T}$ and $x_{\scriptscriptstyle T}$.

If $v(x)$ is even, $i.e.$ $v(-x_{\scriptscriptstyle
T})=v(x_{\scriptscriptstyle T})$ and $v'(-x_{\scriptscriptstyle
T}^{+})=-v'(x_{\scriptscriptstyle T}^{-})$, then
defining the jump of
$v$ at $x$ as $\left[v(x)\right]=v(x^+)-v(x^-)$, the follow equalities
are derived 
\begin{eqnarray} 
\label{eq:evenmatch1}
\left[v(- x_{\scriptscriptstyle T})\right] & = & -\left[v(
x_{\scriptscriptstyle T})\right] \\
\label{eq:evenmatch2}
\left[v'(- x_{\scriptscriptstyle T})\right] & = & \left[v'(
x_{\scriptscriptstyle T})\right] \\
\label{eq:evenmatch3}
\left[v''(- x_{\scriptscriptstyle T})\right] & = & -\left[v''(
x_{\scriptscriptstyle T})\right] \\
\label{eq:evenmatch4}
\left[v'''(- x_{\scriptscriptstyle T})\right] & = & \left[v'''(
x_{\scriptscriptstyle T})\right]
\end{eqnarray}
Given the equalities (\ref{eq:evenmatch1})-(\ref{eq:evenmatch4}), a
direct calculation shows that the matching  
conditions (\ref{eq:L1})-(\ref{eq:L4}) at $-x_{\scriptscriptstyle T}$
are equivalent to 
the matching conditions (\ref{eq:R1})-(\ref{eq:R4}) at
$x_{\scriptscriptstyle T}$.

When $v(x)$ is odd, using the same approach, we can also justify 
that the matching conditions at $-x_{\scriptscriptstyle T}$ and
$x_{\scriptscriptstyle T}$ are the same. 
\end{proof}

\subsection{ODE Solutions}
\label{sec:solutionofodeII}

ODEs I, II, and III are linear with constant coefficients and can be
readily solved in terms of the parameters $A$, $a$, $\alpha$,
and $u_{\scriptscriptstyle T}$.  The eigenvalue $\lambda$
is specified 
when the solutions of the three ODEs are matched across the boundaries
at $x=-x_T$ and $x=x_T$.  Solutions of ODE I are related to ODE III by a reflection $x\rightarrow -x$.  By Lemma
\ref{lem:lem7}, the matching conditions  
at $-x_{\scriptscriptstyle T}$ are the same as those at 
$x_{\scriptscriptstyle T}$.  Thus matching solutions $v_2(x)$  of ODE II
with solutions $v_3(x)$ of ODE III across $x_T$ are sufficient to specify the
eigenvalues of (\ref{eq:eigenvalueequation}).  The solution of  ODE III
is 
$$
v_3(x)=c_5e^{-ax}+c_6e^{-x},
$$
where 
$c_5$ and $c_6$ are constants. Notice that $v_3(x)$ exponentially decays to
zero as $x \rightarrow \infty$, in accordance with the assumed properties of
$v(x)$.

The solutions of ODE II will depend nontrivially on the parameters
$A$, $a$, and $\alpha$.
The characteristic equation of ODE II is
$$
(1+\lambda)\omega^{4}-B \omega^2+C =0,
$$
 where
 \begin{equation}
B=(1+\lambda)(a^2+1)+2\alpha(1-aA)
\label{eq:B}
\end{equation}
and
\begin{equation}
 C=(1+\lambda)a^2-2\alpha a(A-a).
 \label{eq:C}
 \end{equation}
The characteristic values are
\begin{equation}
\omega^2=\frac{B\pm \sqrt{\Delta}}{2(1+\lambda)}
\label{charvals}
\end{equation}
where
\begin{eqnarray}
\label{eq:delta}
\Delta  & = & B^2-4(1+\lambda)C \\
& = & (a^2-1)^2
\lambda^2+2(a^2-1)(a^2-1-2aA\alpha-2\alpha)\lambda- \nonumber\\ & &
(a^2-1)(1-a^2+4\alpha+4aA\alpha)+4 \alpha^2(1-aA)^2\nonumber
\end{eqnarray}

Let $\lambda_B$ be the zero of $B$. If $\Delta$ 
is negative, (\ref{charvals}) shows that ODE II will have complex characteristic values.  If  $\Delta$ is positive,  combinations of $B$ and $\Delta$ yield either real or imaginary values.
For fixed
$A$, $a$ and $\alpha$, $\Delta$ is a parabola with a left zero
$\lambda_l$ and a right zero $\lambda_r$.  By Lemma (\ref{lem:lem8}) and Lemma (\ref{lem:lem9})  in the Appendix,   either $\lambda_l \le \lambda_B \le \lambda_r$ and does not intersect with either branch of $\sqrt{\Delta}$ or
 $\lambda_B\le \lambda_l$ and intersects with the left branch of $\sqrt{\Delta}.$ Table 
\ref{tab:characteristic1} and \ref{tab:characteristic2} describe all the 
possible structures of the characteristic values $\pm\omega_1$ and
$\pm \omega_2$.
There are three possible forms of solution $v_2(x)$: 1) both $\omega_1$ and $\omega_2$ are real; 2) both $\omega_1$ and $\omega_2$ are complex; 3) $\omega_1$ is real and $\omega_2$ is imaginary.
\begin{table}[!hbt]
\centering
\caption{Characteristic value chart when $\lambda_l<\lambda_B<\lambda_r$}
\begin{small}
\begin{tabular}{|c|c|c|c|c|c|} 
\hline
   1  & 2 & 3 & 4& 5 \\ \hline
     $-1<\lambda<\lambda_l$ & $\lambda=\lambda_l$ & $\lambda_l<\lambda<\lambda_r$ & $\lambda=\lambda_r$ & $\lambda>\lambda_r$\\ \hline
   $B<0$ &$B<0$ &$B>0$ or $B<0$ &$B>0$ & $B>0$ \\ \hline 
 $\Delta>0$, & $\Delta=0$ & $\Delta<0$ & $\Delta=0$ & $\Delta>0$\\
$|B|<\sqrt{\Delta}$ &  &  & & \\ \hline 
  $\omega_1$ real & $\omega_{1,2}$ imaginary & $\omega_{1,2}$ complex & $\omega_{1,2}$ real &$\omega_{1,2}$ real  \\
 $\omega_2$ imaginary & $\omega_1=\omega_2^*$ & $\omega_1=\omega_2^*$ & $\omega_1=\omega_2$ &  \\ \hline
\end{tabular}
\end{small}
\label{tab:characteristic1}
\end{table}
\begin{table}[!hbt]
\centering
\caption{Characteristic value chart when $\lambda_B<\lambda_l<\lambda_r$}
\begin{small}
\begin{tabular}{|c|c|c|c|c|c|c|c|} \hline
   1  & 2 & 3 & 4& 5& 6 \\ \hline
     $-1<\lambda<\lambda_I$ & $\lambda_I \leq \lambda<\lambda_l$ & $\lambda=\lambda_l$ & $\lambda_l<\lambda<\lambda_r$ & $\lambda=\lambda_r$ & $\lambda>\lambda_r$\\ \hline
   $B<0$  or & $B>0$& $B>0$ &$B<0$  &$B>0$ & $B>0$ \\
 $B>0$ &  &  &  &  & \\ \hline 
 $\Delta>0$,  &$\Delta>0$,& $\Delta=0$ & $\Delta<0$ & $\Delta=0$ & $\Delta>0$\\
 $|B|<\sqrt{\Delta}$ & $|B|>\sqrt{\Delta}$ &  &  & & \\ \hline 
  $\omega_1$ real &$\omega_{1,2}$ real & $\omega_{1,2}$ real & $\omega_{1,2}$ complex & $\omega_{1,2}$ real & $\omega_{1,2}$ real \\
 $\omega_2$ imaginary & & $\omega_1=\omega_2$ & $\omega_1=\omega_2^*$ & $\omega_1=\omega_2$ &  \\ \hline
\end{tabular}
\end{small}
\label{tab:characteristic2}
\end{table}

We notate the even solutions of ODE II as 
$v_2^\mathrm{e}(x)$ and the odd solutions  as $v_2^\mathrm{o}(x)$.
When $\lambda \geq \lambda_r$ or $\lambda_I \leq \lambda \leq \lambda_l $, 
both $\omega_1$ and $\omega_2$ are real.  Thus
\begin{eqnarray}
\label{eq:realevenv2}
v_2^\mathrm{e}(x)=c_3 \mu_1(x)+c_4\frac{\mu_1(x)-\mu_2(x)}{\omega_1-\omega_2}
\end{eqnarray}
where $\mu_1(x)=e^{\omega_1 x}+e^{-\omega_1 x},$ and $\mu_2(x)=e^{\omega_2 x}+e^{-\omega_2 x}$.  We use (\ref{eq:realevenv2}) because it is more convenient to resolve the degenerate case of $\omega_1=\omega_2$.
As $\lambda \rightarrow  \lambda_r^-,$ $\mu_1 \rightarrow \mu_2$ and $\epsilon=\omega_1-\omega_2 \rightarrow 0,$ (\ref{eq:realevenv2}) becomes
\begin{eqnarray*}
v_2^\mathrm{e}(x) =
& = &  c_3(e^{\omega_1 x}+e^{-\omega_1 x})+c_4\frac{(e^{\omega_1 x}+e^{-\omega_1 x})-(e^{\omega_1x}e^{-\epsilon x}+e^{-\omega_1 x}e^{\epsilon x})}{\epsilon}
\end{eqnarray*}
Replacing $e^{\epsilon x}$ by $1+\epsilon x$, $e^{-\epsilon x}$ by $1-\epsilon x$ and taking the limit
as $\epsilon \rightarrow 0$ yields
\begin{eqnarray}
v_2^\mathrm{e}(x) & = &  c_3(e^{\omega_1 x}+e^{-\omega_1 x})+c_4x (e^{\omega_1 x}-e^{-\omega_1 x})\nonumber\\
\label{eq:realmatchcomp}
& = & 2c_3 \cosh px + 2 c_4 x \sinh px 
\end{eqnarray}
(\ref{eq:realevenv2}) approaches (\ref{eq:realmatchcomp})
as $\lambda \rightarrow  \lambda_r^-.$ It matches the solution 
${\displaystyle v_2^\mathrm{e}(x)}$ as $\lambda \rightarrow  \lambda_r^+,$ 
which is given in (\ref{eq:compv2e}). 
 
Similarly, $v_2^\mathrm{o}(x)$ can be written as 
\begin{eqnarray*}
v_2^\mathrm{o}(x)=c_3(e^{\omega_1 x}-e^{-\omega_1 x})+c_4\frac{(e^{\omega_1 x}-e^{-\omega_1 x})-(e^{\omega_2 x}-e^{-\omega_2 x})}{\omega_1-\omega_2}
\end{eqnarray*}

When $\lambda_l<\lambda<\lambda_r$,
$\omega_1$ and $\omega_2$ are complex. Let $\omega_1=p+iq$, $\omega_2=p-iq$. When $v_2(x)$ is even,  write $v_2^\mathrm{e}(x)$ as 
\begin{eqnarray}
v_2^\mathrm{e}(x)=2 c_3 \cos qx \cosh px +2 c_4 \frac{\sin qx}{q} \sinh px
\end{eqnarray}
As $\lambda \rightarrow \lambda_l^+$ or $\lambda_r^-$, $q \rightarrow 0$,
\begin{eqnarray}
\label{eq:compv2e}
v_2^\mathrm{e}(x) \rightarrow 2 c_3 \cosh px +2 c_4 x \sinh px
\end{eqnarray}
$v_2^\mathrm{o}(x)$ can be written as 
\begin{eqnarray*}
v_2^\mathrm{o}(x) = 2 c_3 \cos qx \sinh px -2 c_4 \frac{ \sin qx}{q} \cosh px
\end{eqnarray*}
where ${\displaystyle
  p=\sqrt{\frac{\sqrt{B^2+|\Delta|}}{2(1+\lambda)}}}\cos \theta$,
${\displaystyle p=\sqrt{\frac{\sqrt{B^2+|\Delta|}}{2(1+\lambda)}}}\sin
\theta$ and ${\displaystyle \theta=\frac{1}{2}\arctan
  \frac{\sqrt{|\Delta|}}{B}}$.

When $-1<\lambda<\lambda_I$, $\omega_1$ is real and $w_2$ is imaginary. Let 
$\omega_2=iq$, where ${\displaystyle q=\sqrt{\frac{\sqrt{\Delta}-B}{2(1+\lambda)}}}$, then 
\begin{eqnarray}
v_2^\mathrm{e}(x) & = &  c_3(e^{\omega_1 x}+e^{-\omega_1 x})+2 c_4\cos(qx) \\
\label{eq:realimageven}
v_2^\mathrm{o}(x) & = &  c_3(e^{\omega_1 x}-e^{-\omega_1 x})+2 c_4\frac{\sin(qx)}{q}
\label{eq:realimagodd}
\end{eqnarray}

\section{Stability criteria}
\label{sec:stabilitycriteria}
By theorem \ref{the:the3}, $v_1(x)$ and $v_2(x)$ must match at
$-x_{\scriptscriptstyle T}$, and $v_2(x)$  and $v_3(x)$ must match at
$x_{\scriptscriptstyle T}$. By  
property \ref{lem:lem7}, the matching  
conditions at $-x_{\scriptscriptstyle T}$ are same as the matching conditions 
at $x_{\scriptscriptstyle T}$ for $v(x)$ even or odd. Therefore, it
suffices to apply the matching  
condition to $v_2(x)$ and $v_3(x)$ at $x_{\scriptscriptstyle T}$ for the even 
and odd cases separately.  This reduces the dimensionality of the
resulting eigenvalue problem by a factor of two. 
In general, the matching conditions can be written 
as 
\begin{eqnarray*}
T1: \hspace{0.2cm} \left \{ \begin{array}{lcl}
\left[v(x_{\scriptscriptstyle T})\right]=v_3(x_{\scriptscriptstyle T}^+)-v_2(x_{\scriptscriptstyle T}^-) & = & 0\\
\left[v'(x_{\scriptscriptstyle T})\right]=v_3'(x_{\scriptscriptstyle T}^+)-v_2'(x_{\scriptscriptstyle T}^-)& = &
{\displaystyle \frac{2 \alpha (1-a A)}{1+\lambda}v( x_{\scriptscriptstyle T})}\\
\left[v''(x_{\scriptscriptstyle T})\right]=v_3''(x_{\scriptscriptstyle T}^+)-v_2''(x_{\scriptscriptstyle T}^-) & = &{\displaystyle
\frac{ 2(a A-1)}{c(1+\lambda)}v( x_{\scriptscriptstyle T})}\\
\left[v'''(x_{\scriptscriptstyle T})\right]=v_3'''(x_{\scriptscriptstyle T}^+)-v_2'''(x_{\scriptscriptstyle T}^-)& = &{\displaystyle
\frac{2(1-a^3A)}{c(1+\lambda)}v( x_{\scriptscriptstyle T})+
\frac{2 \alpha (aA-1)}{1+\lambda}v'( x_{\scriptscriptstyle T}^{-})}
\end{array}\right.
\end{eqnarray*}
where $v( x_{\scriptscriptstyle T})=v_3(x_{\scriptscriptstyle T}^+) $ and  $v'( x_{\scriptscriptstyle T}^{-})=v_2'(x_{\scriptscriptstyle T}^-)$.

 A given stationary pulse solution $u_0(x)$ will be specified by a set of parameters
 $a$, $A$, $\alpha$, $x_{\scriptscriptstyle T}$,  and $u_{\scriptscriptstyle T}$.
The eigenvalues $\lambda$ that determine stability of pulse solutions are given by system T1.   To compute these eigenvalues, we require the appropriate form of the eigenfunctions $v_2(x)$ and $v_3(x)$.  We do so by
finding characteristic values (\ref{charvals}) corresponding to the parameters specifying the given stationary pulse solution.   We expediate this process by calculating the constants $B$ (\ref{eq:B}) and $C$ (\ref{eq:C}), then using Table~\ref{tab:characteristic1} and Table~\ref{tab:characteristic2} to deduce the characteristic value types.  We then substitute the appropriate form for $v_2(x)$ and $v_3(x)$ into $T_1$ where coefficients $c_3$ and $c_4$ in
$v_2(x_{\scriptscriptstyle T})$ and $c_5$ and $c_6$ in $v_3(x_{\scriptscriptstyle T})$ are unknown. We replace
$v(x_{\scriptscriptstyle T})$ by $v_3(x_{\scriptscriptstyle T}^+)$ and
$v'(x_{\scriptscriptstyle T}^-)$ by $v_2'(x_{\scriptscriptstyle T}^-)$. 
This results in a $4 \times 4$ homogeneous
linear system with $4$ unknown free parameters $c_3$, $c_4$, $c_5$,
$c_6$.  We must do this for both even and odd eigenfunctions resulting in two separate linear systems that must be solved. 

The coefficient matrix of this system must be singular for a non-trivial solution $(c_3, c_4, c_5, c_6)$.  Hence, the determinant $D(\lambda)$ of the coefficient
matrix must be zero. Thus, the solution
of $D(\lambda)=0$ is an eigenvalue and it determines the stability
of the stationary solution.  If there exists a $\lambda$ such that
$0<\lambda<\lambda_b$ and $D(\lambda)=0,$ then the standing pulse is
unstable. If there is no positive $\lambda$ such that
$0<\lambda<\lambda_b$ and $D(\lambda)=0,$ the standing pulse is
stable.  Our determinant $D(\lambda)$ for stability is similar to the Evans Function~\cite{Evans, E1, E2, E3}.

\subsection{Stability of the small and large pulse}
\label{sec:examples}
Two single-pulse solutions were shown to exist in the accompanying paper \cite{Guo1} for parameters $a=2.4$, $A=2.8$, $\alpha=0.22$, $u_{\scriptscriptstyle T}=0.400273$ and $\beta=1$.  The large pulse has a higher amplitude and larger width and is denoted by $u^{\bf l}(x)$. The small pulse   is slightly 
above threshold and much narrower than $u^{\bf l}(x)$ and is denoted by $u^{\bf s}(x)$. The explicit forms are given by 
\begin{small}
\begin{eqnarray*}
u^{\bf l}(x)=\left \{ \begin{array}{ll}
0.665 \cos (0.31 x) \cosh (1.49x)-3.78 \sin(0.31x) \sinh (1.49x)+0.33,  x \in [-x_{\scriptscriptstyle T}, x_{\scriptscriptstyle T}]
 \\
6.237 e^{-2.4 |x|}-1.604 e^{-|x|},  \mbox {\hspace{6cm} otherwise}
\end{array} \right. \end{eqnarray*}
\end{small}
where $x_{\scriptscriptstyle T}=0.683035$.
\begin{small}
\begin{eqnarray*}
u^{\bf s}(x)=\left \{ \begin{array}{ll}
0.22 \cos (0.31 x) \cosh (1.49x)-8.03\sin(0.31x) \sinh (1.49x)+0.33, x \in [-x_{\scriptscriptstyle T}, x_{\scriptscriptstyle T}]
 \\
1.203 e^{-2.4 |x|}-0.416 e^{-|x|}, \mbox{\hspace{6cm} otherwise}
\end{array} \right.
\end{eqnarray*}
\end{small}
where $x_{\scriptscriptstyle T}=0.202447$.

We first calculate the upper bound for the eigenvalue $\lambda_b$  which is different for the large pulse and small pulse because $\lambda_b$ depends 
on $x_{\scriptscriptstyle T}$. Let $\lambda_b^{\bf l}$ be the upper bound for the large pulse and $\lambda_b^{\bf s}$ be the upper bound for 
the small pulse. For the parameter set  $a=2.4,
A=2.8,\alpha=0.22,u_{\scriptscriptstyle T}=0.400273$, the upper bounds are
$\lambda_b^{\bf l}=1.25917$ and $\lambda_b^{\bf s}=1.66628.$

For the above set of parameters,
$v_3(x)$ always has the following form. 
\begin{eqnarray*}
v_3(x)=c_5e^{-ax}+c_6e^{-x}.
\end{eqnarray*}
The form of $v_2(x)$ depends on $\omega_1$ and $\omega_2$. For this specific set of parameters, the left and right solutions of $\Delta$ (\ref{eq:delta}) are $\lambda_l=-0.627692$ and $\lambda_r=0.192861$. 
When $0\leq \lambda \leq \lambda_r$, both $\omega_1$ and $\omega_2$ are complex, implying
\begin{eqnarray*}
 v_2(x)= \left \{ \begin{array}{ll}
{\displaystyle v_2^\mathrm {e}(x)= 2 c_3 \cos qx \cosh px +2 c_4\frac{\sin q x}{q} \sinh px}  \mbox{\hspace{1cm} $v_2(x)$ is even}\\
{\displaystyle v_2^\mathrm {o}(x) =2 c_3 \cos qx \sinh px -2 c_4 \frac{ \sin qx}{q}} \cosh px \mbox{\hspace{1cm} $v_2(x)$ is odd}
\end{array} \right.
\end{eqnarray*}
where $p,$ $q$ are real, and $c_3$, $c_4$ are unknown.
 
Substituting $v_2^\mathrm{e}(x)$ ($v_2^\mathrm{o}(x)$) and $v_3(x)$ into system $T1$, results in an unwieldy $4 \times 4$ linear system in $c_3$, $c_4$, $c_5$ and $c_6$. 
We use Mathematica \cite{Mathematica} to calculate 
the determinant of the coefficient matrix as a 
function of $\lambda$.  

\begin{figure} [!htb]  
\centering
\includegraphics[height=2.5in]{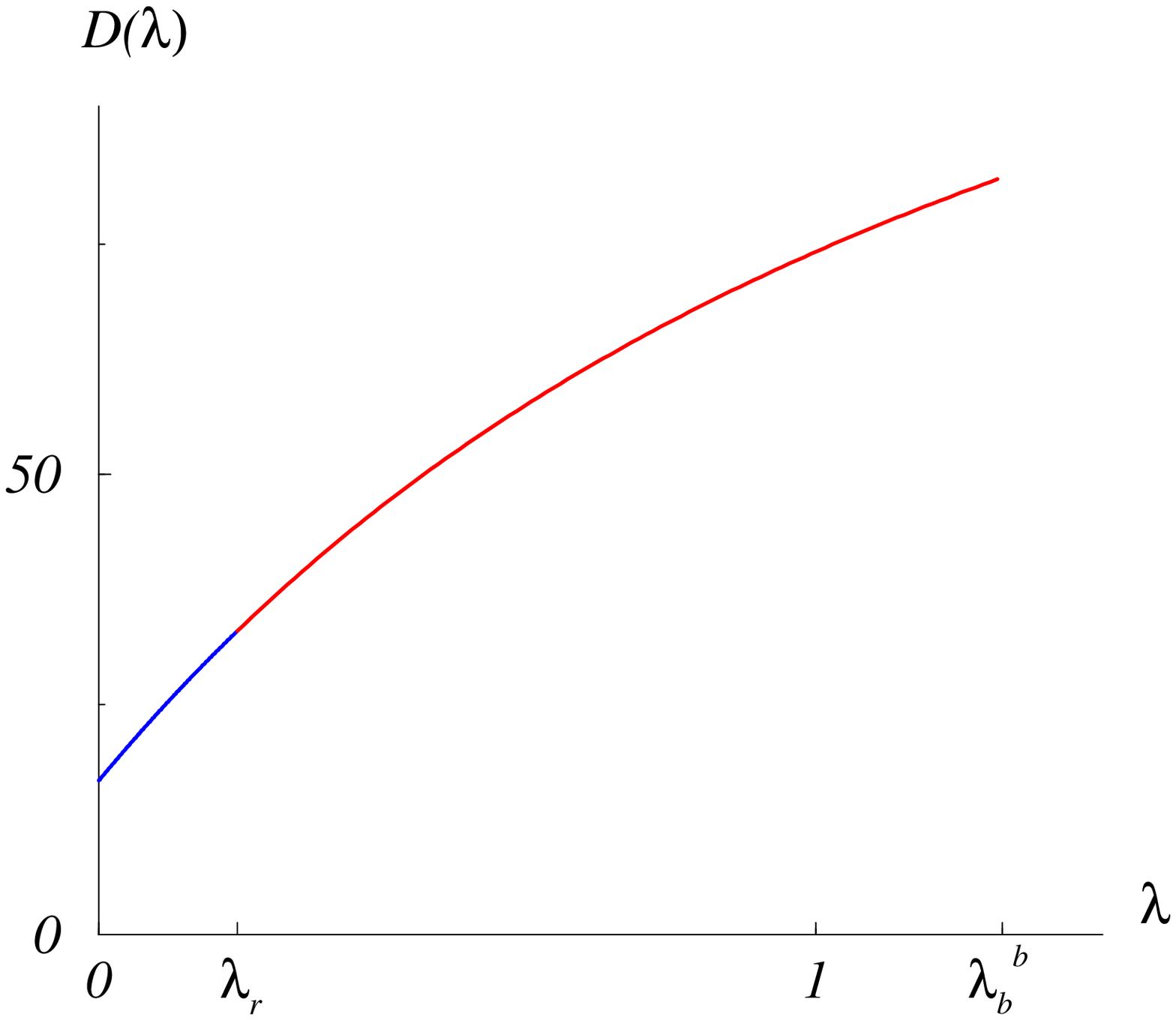}
\caption{\small Plot of $D(\lambda)$ for large single-pulse $u^{\bf l}(x)$ 
when $v_2(x)$ is even.
$a=2.4$, $A=2.8$, $\alpha=0.22$, $u_{\scriptscriptstyle T}=0.400273$, $x_{\scriptscriptstyle T}=0.683035$, $\lambda_r=0.192861$, $\lambda_b^{\bf l}=1.25917$. There is no positive $\lambda$ such that $D(\lambda)=0$, $\lambda \leq \lambda_b^{\bf l}$.}
\label{fig:evenbigbump} 
\end{figure}

\begin{figure} [!htb]
\centering
\includegraphics[height=2.5in]{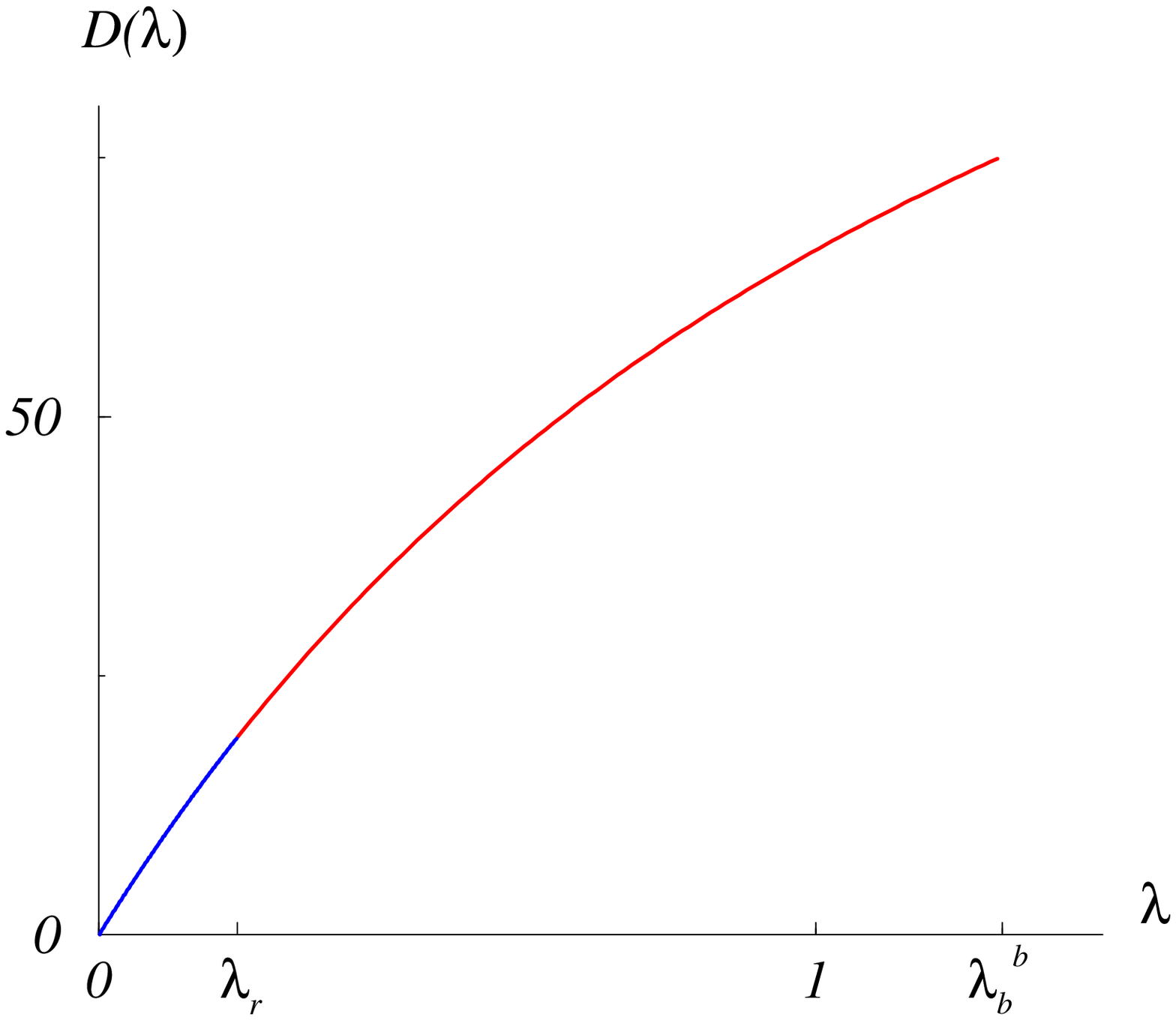}
\caption{\small Plot of $D(\lambda)$ for large single-pulse $u^{\bf l}(x)$ 
when $v_2(x)$ is odd. $a=2.4$, $A=2.8$, $\alpha=0.22$, $u_{\scriptscriptstyle T}=0.400273$, $x_{\scriptscriptstyle T}=0.683035$, $\lambda_r=0.192861$, $\lambda_b^{\bf l}=1.25917$. There is no positive  $\lambda$ such that $D(\lambda)=0$, 
$\lambda \leq \lambda_b^{\bf l}.$ When $v_2(x)$ is odd, $D(\lambda)$ does 
identify the zero eigenvalue.}
\label{fig:oddbigbump} 
\end{figure}

When $0.192861=\lambda_r \leq \lambda \leq \lambda_b^{\bf l}=1.25917$, 
$\omega_{1,2}$ is real, and
$v_2(x)$ has the form
\begin{eqnarray*}
v_2(x)=\left \{\begin{array}{ll}
{\displaystyle c_3(e^{\omega_1 x}+e^{-\omega_1 x})+c_4\frac{(e^{\omega_1 x}+e^{-\omega_1 x})-(e^{\omega_2 x}+e^{-\omega_2 x})}{\omega_1-\omega_2}} \mbox{\hspace{0.6cm}  $v_2(x)$ is even} \\
{\displaystyle c_3(e^{\omega_1 x}-e^{-\omega_1 x})-c_4\frac{(e^{\omega_1 x}-e^{-\omega_1 x})-(e^{\omega_2 x}+e^{-\omega_2 x})}{\omega_1-\omega_2}} \mbox{\hspace{0.6cm} $v_2(x)$ is odd} \end{array} \right.
\end{eqnarray*}
Figure~\ref{fig:evenbigbump} gives a plot of $D(\lambda)$ on the domain $[0,\lambda_b]$,  combining the regimes where  $\omega_{1,2}$ is real and complex.   We see that there is no positive $\lambda$ that satisfies $D(\lambda)=0$. 
Figure~\ref{fig:oddbigbump} shows $D(\lambda)$ for odd $v(x)$.  
We see that $D(\lambda)=0$ only when $\lambda=0$, which is consistent with Theorem \ref{the:the9}.  The lack of a positive zero of $D(\lambda)$ indicates that the large pulse  is stable.

\begin{figure}[!htb]
\centering
\epsfig{figure=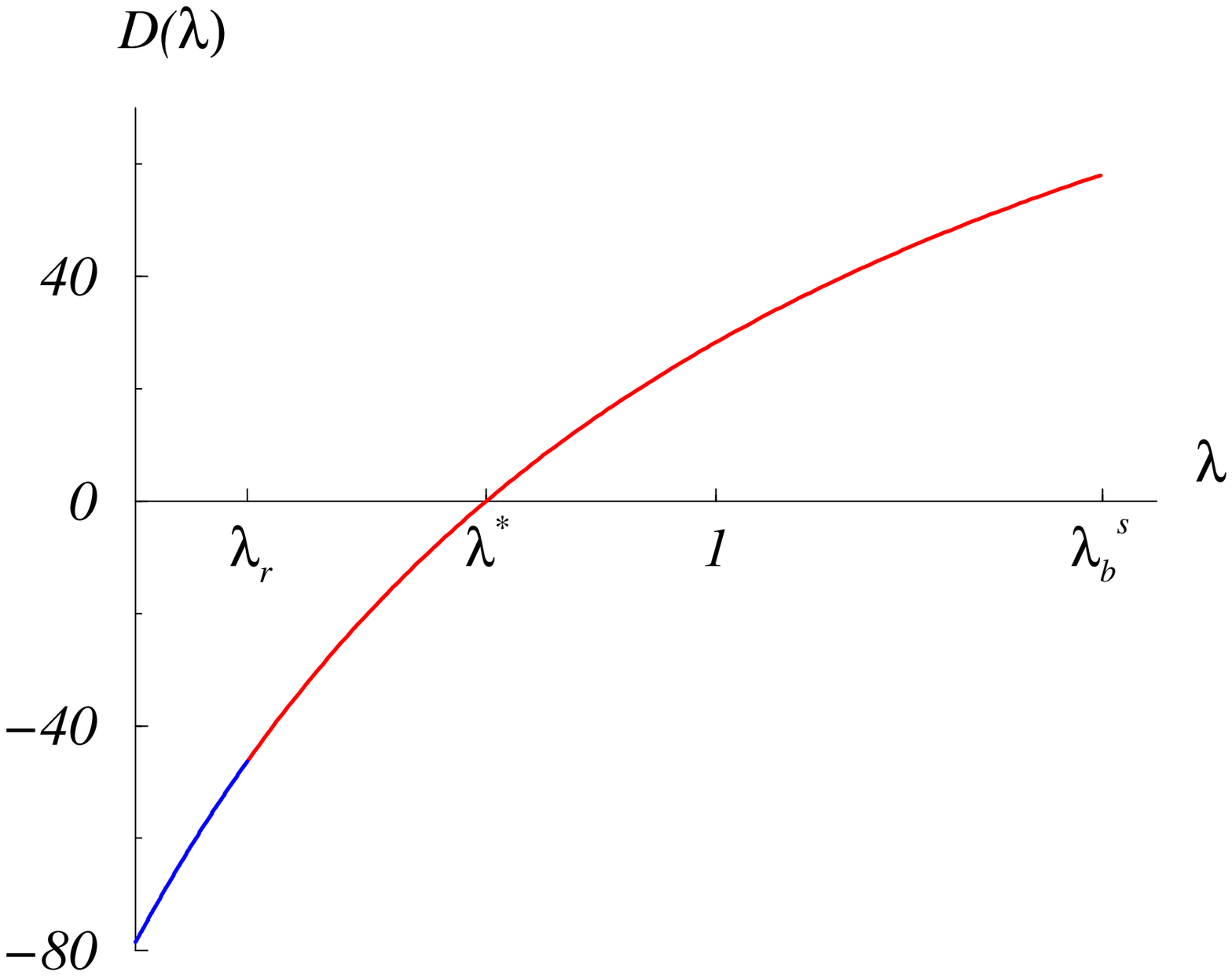, height=2.5in}
\caption{\small Plot of $D(\lambda)$ for small single-pulse $u^{\bf s}(x)$when $v_2(x)$ is even.
$a=2.4$, $A=2.8$, $\alpha=0.22$, $u_{\scriptscriptstyle T}=0.400273$, $x_{\scriptscriptstyle T}=0.683035$, $\lambda_r=0.192861$, $\lambda_b^{\bf s}=1.66628$. $\lambda^*=0.603705$ There is one positive $\lambda=\lambda^*$ such that $D(\lambda^*)=0$, $\lambda^* \leq \lambda_b^{\bf s}$.}
\label{fig:evensmalbump}
\end{figure}

\begin{figure}
\centering
\epsfig{figure=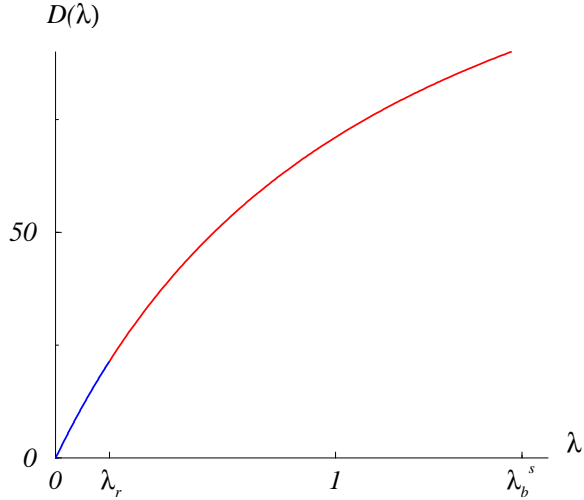, height=2.5in}
\caption{\small Plot of $D(\lambda)$ for small single-pulse $u^{\bf s}(x)$ 
when $v_2(x)$ is odd.   
$a=2.4$, $A=2.8$, $\alpha=0.22$, $u_{\scriptscriptstyle T}=0.400273$, $x_{\scriptscriptstyle T}=0.683035$, $\lambda_r=0.192861$, $\lambda_b^{\bf s}=1.66628$. There is no positive  $\lambda$ such that $D(\lambda)=0$, $\lambda \leq \lambda_b^{\bf s}.$ When $v_2(x)$ is odd, $D(\lambda)=0$ at $\lambda=0$ identifies the zero eigenvalue.}
\label{fig:oddsmalbump}
\end{figure}

 
For the same set of parameters, $\left \{ a=2.4, A=2.8, \alpha=0.22,
u_{\scriptscriptstyle T}=0.400273\right \}$, the upper bound of the
small pulse is $\lambda_b^{\bf s}=1.66628$  
Repeating the same procedure as for the large pulse, we plot $D(\lambda)$ for 
both $v_2^\mathrm{e}(x)$ and $v_2^\mathrm{o}(x)$
(Figs~\ref{fig:evensmalbump} and \ref{fig:oddsmalbump}). 
The existence of a positive eigenvalue $\lambda=\lambda^*$ satisfying
$D(\lambda^*)=0$ in  
Fig.~\ref{fig:evensmalbump} implies the instability of the small single-pulse. 
The plot of $D(\lambda)$ corresponding to $v^\mathrm{o}(x)$ in figure
\ref{fig:oddsmalbump} identifies the zero eigenvalue.


\subsection{Stability and instability of for different gain $\alpha$}
\label{sec:stabilityforrange}
\begin{figure}[!htb]
\centering
\epsfig{figure=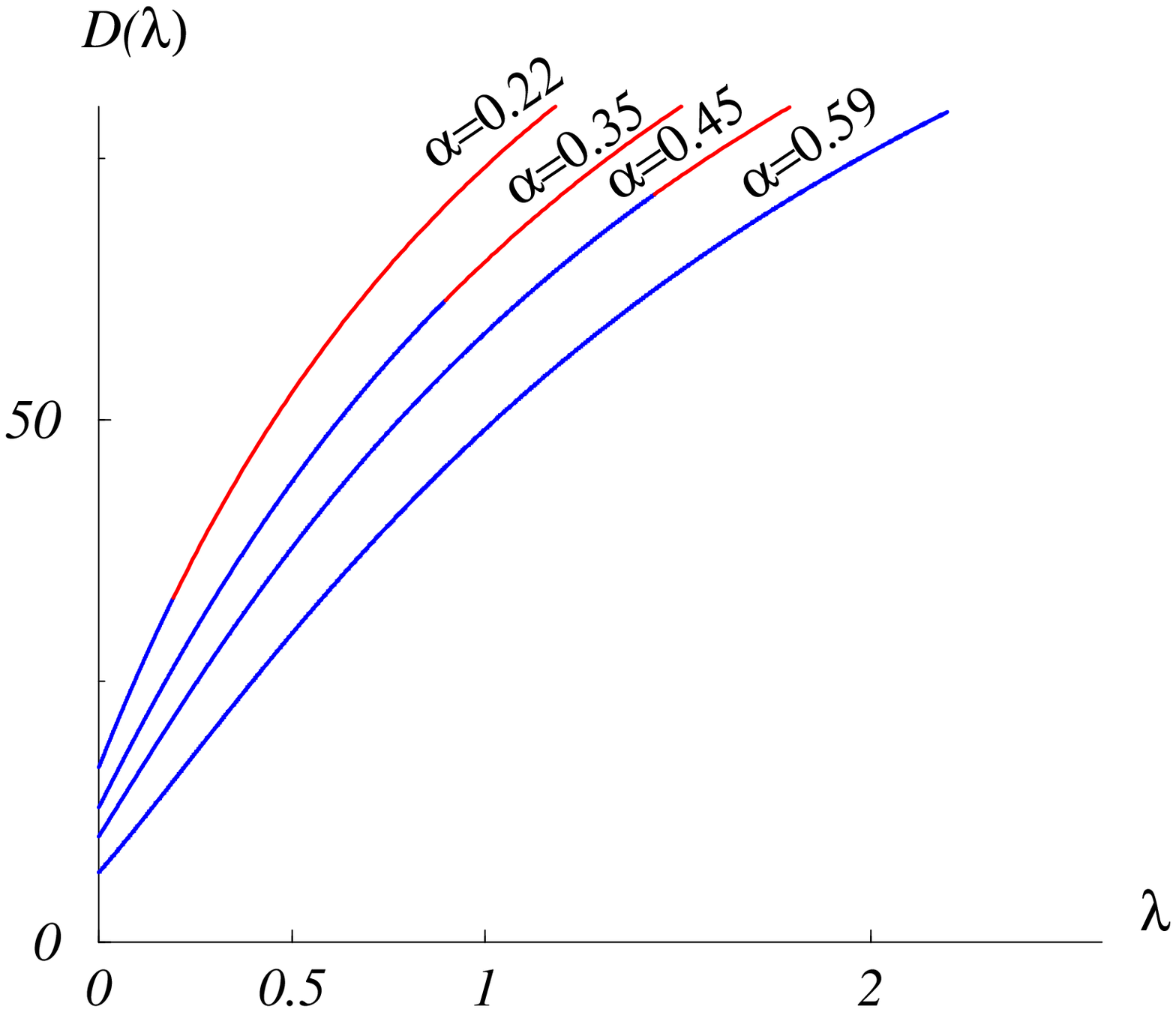,height=2.5in}
\caption{\small Plots of $D(\lambda)$ for large single-pulses with different
gain $\alpha.$ $a=2.4$, $A=2.8$, $\alpha=0.22$, $u_{\scriptscriptstyle T}=0.400273$.}
\label{fig:stabilitydetplots}
\end{figure}

\begin{figure}[!htb]
\centering
\epsfig{figure=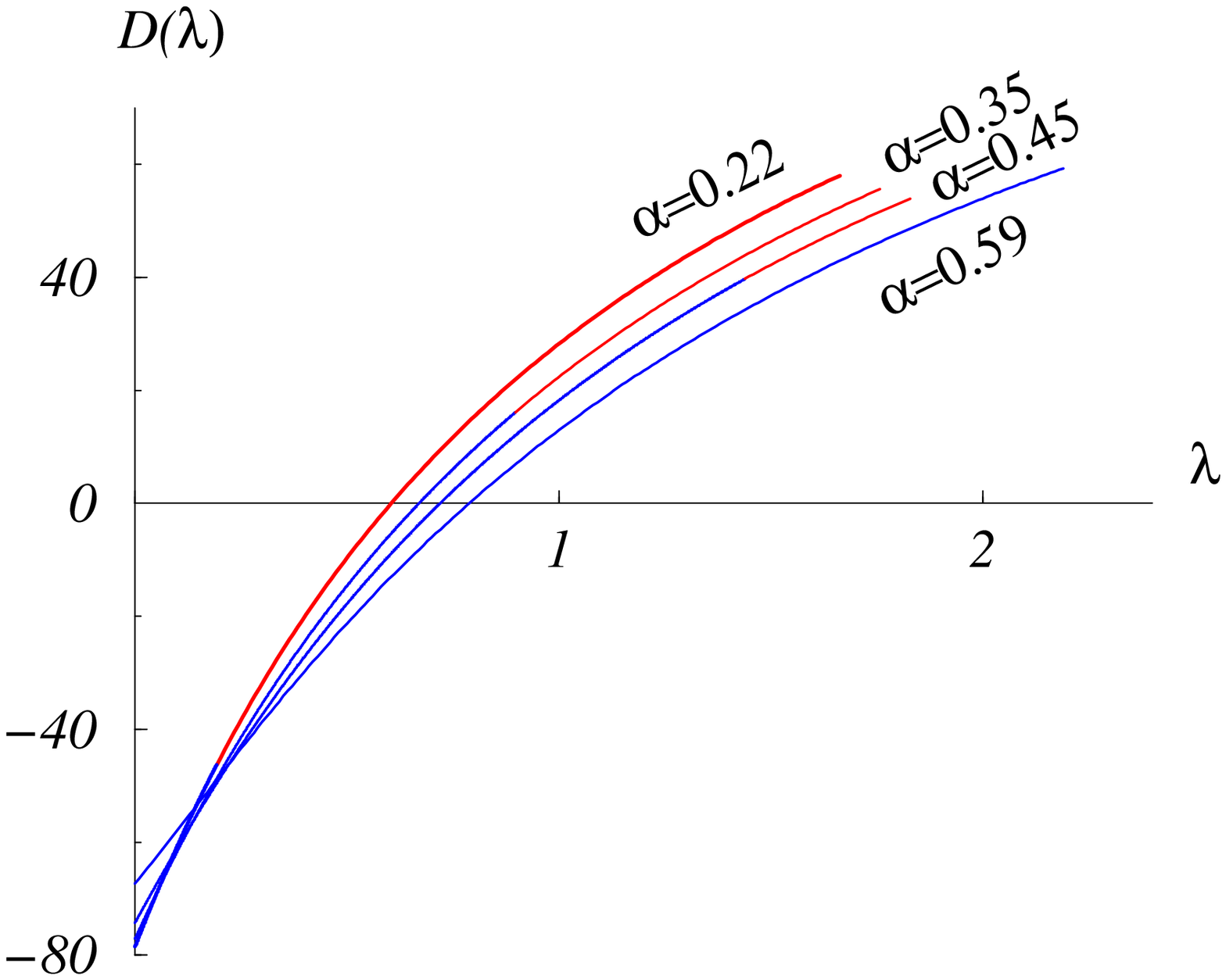, height=2.5in}
\caption{\small  Plots of $D(\lambda)$ for small single-pulses with different
gain $\alpha.$ $a=2.4$, $A=2.8$, $\alpha=0.22$,  $u_{\scriptscriptstyle T}=0.400273$.}
\label{fig:stabidetsmalplots}
\end{figure}

For both the large single-pulses and small single-pulses, $D(\lambda)$ is 
monotonically increasing (See Fig \ref{fig:stabilitydetplots} and 
\ref{fig:stabidetsmalplots}).
\begin{figure}
\centering
\epsfig{figure=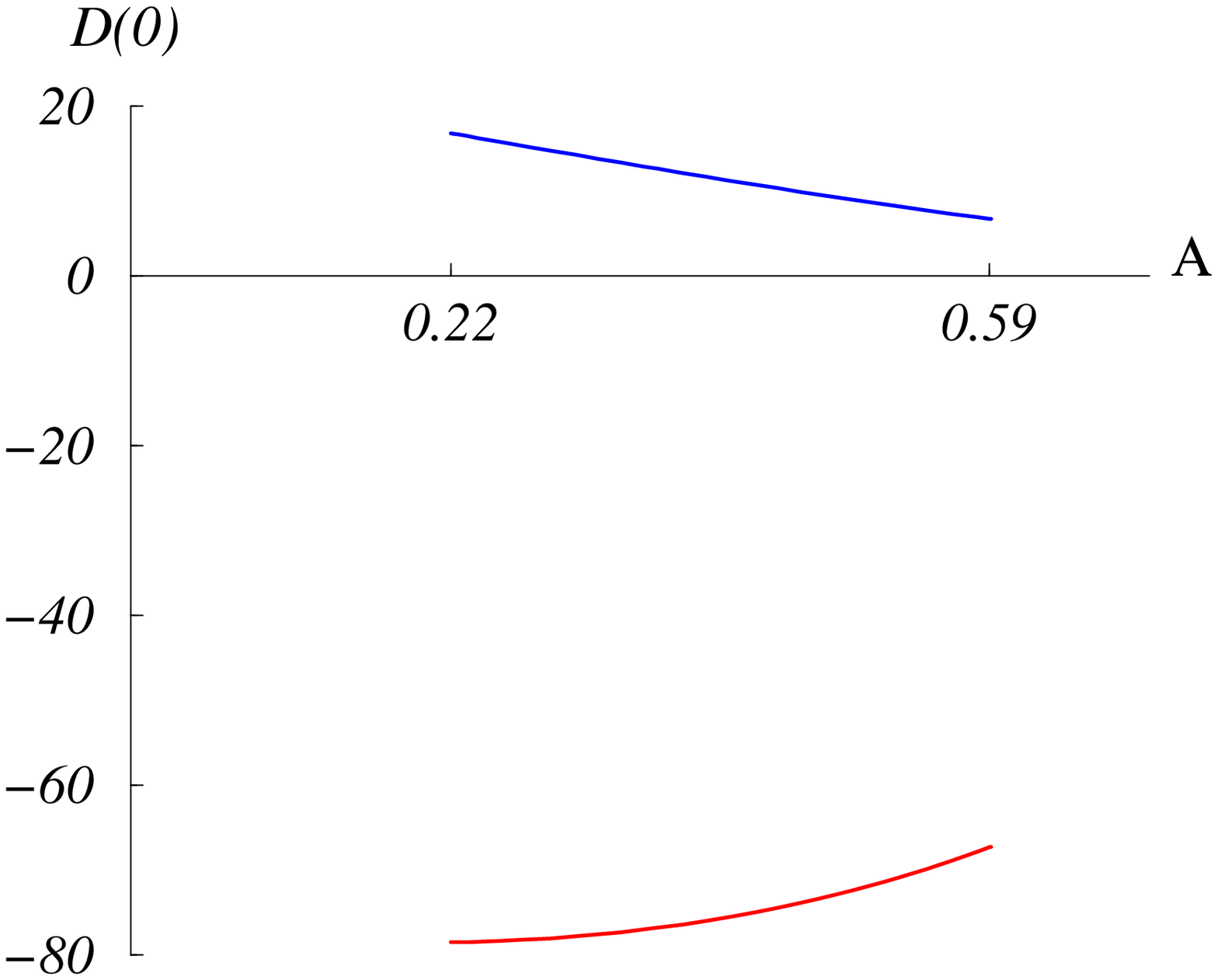, height=2.5in}
\caption{\small Plots of $D(0)$ for both large single-pulses (blue branch) 
and small single-pulse (red branch) with $\alpha
 \in (0.22, 0.59).$ $a=2.4,$ $A=2.8.$,  $u_{\scriptscriptstyle T}=0.400273$.}
\label{fig:d0plots}
\end{figure}
However, $D(0)$ for small pulses is negative. As 
$\lambda$ increases, $D(\lambda)$ crosses the $\lambda$-axis and
becomes positive.  
Therefore, $D(\lambda)$ has a positive zero. For the large pulse, $D(0)$ is 
positive and $D(\lambda)$ has no positive zero. We follow $D(0)$ for a
range of  
$\alpha \in (0.22, 0.59)$ in Fig \ref{fig:d0plots} and find that
$D(0)$ is always negative for small pulses and positive for large 
pulses.  Hence, the large pulses are stable and the small pulses are unstable 
in this range. 

\subsection{ Stability of the dimple-pulse $u^{\bf d}(x)$  and the instability of the third pulse}
\begin{figure}[!htb] 
\centering
\epsfig{figure=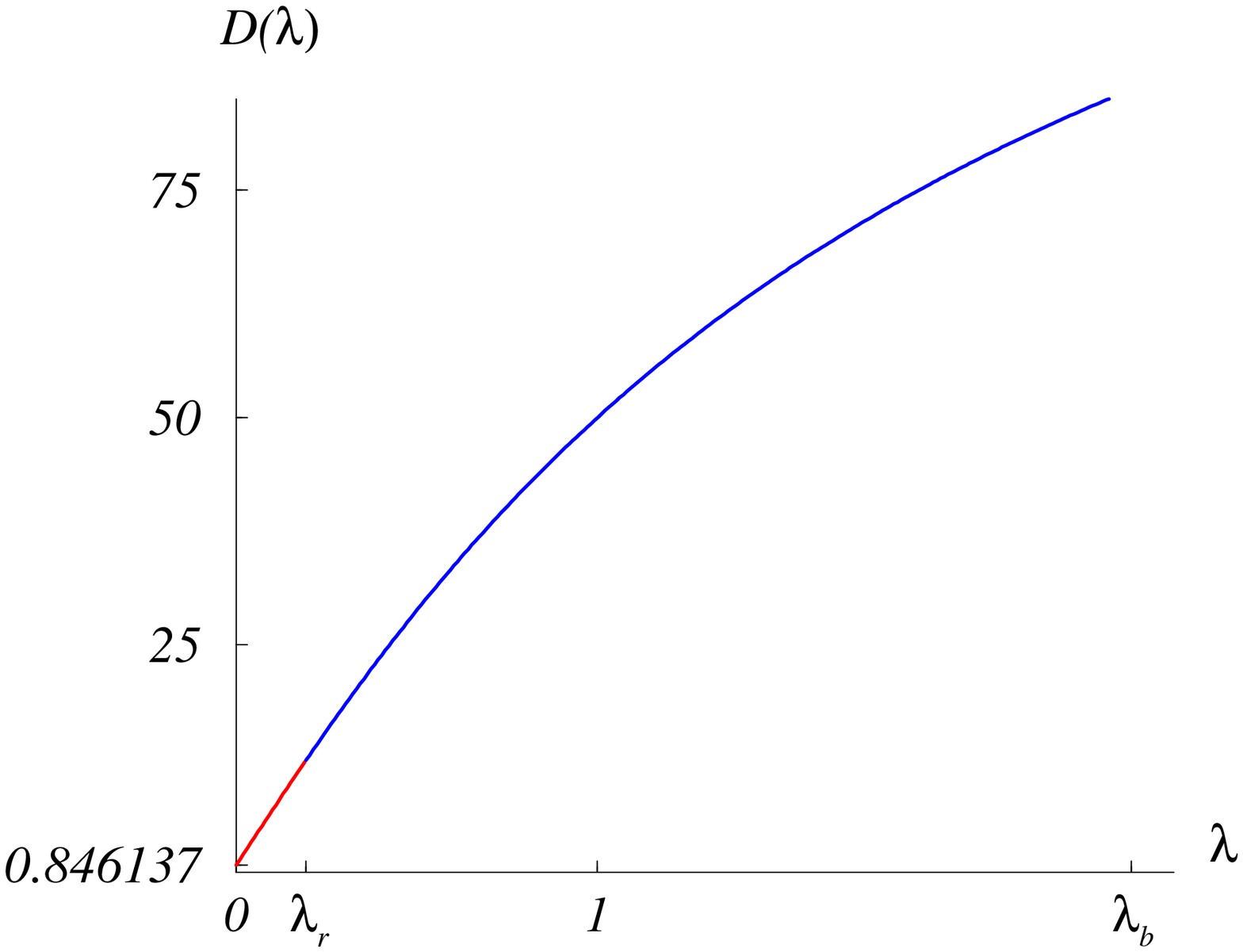, height=2.5in}
\caption{\small Plot of $D(\lambda)$ for dimple-pulse when $v_2(x)$ is even
$a=2.4$, $A=2.8$, $\alpha=0.22$, $x_{\scriptscriptstyle T}=2.048246$, $\lambda_r=0.192861$, $\lambda_b^{\bf d}=2.48147$. There is no positive $\lambda$ such that $D(\lambda)=0$.}
\label{fig:evendimpledet}
\end{figure}

\begin{figure}[!htb] 
\centering
\epsfig{figure=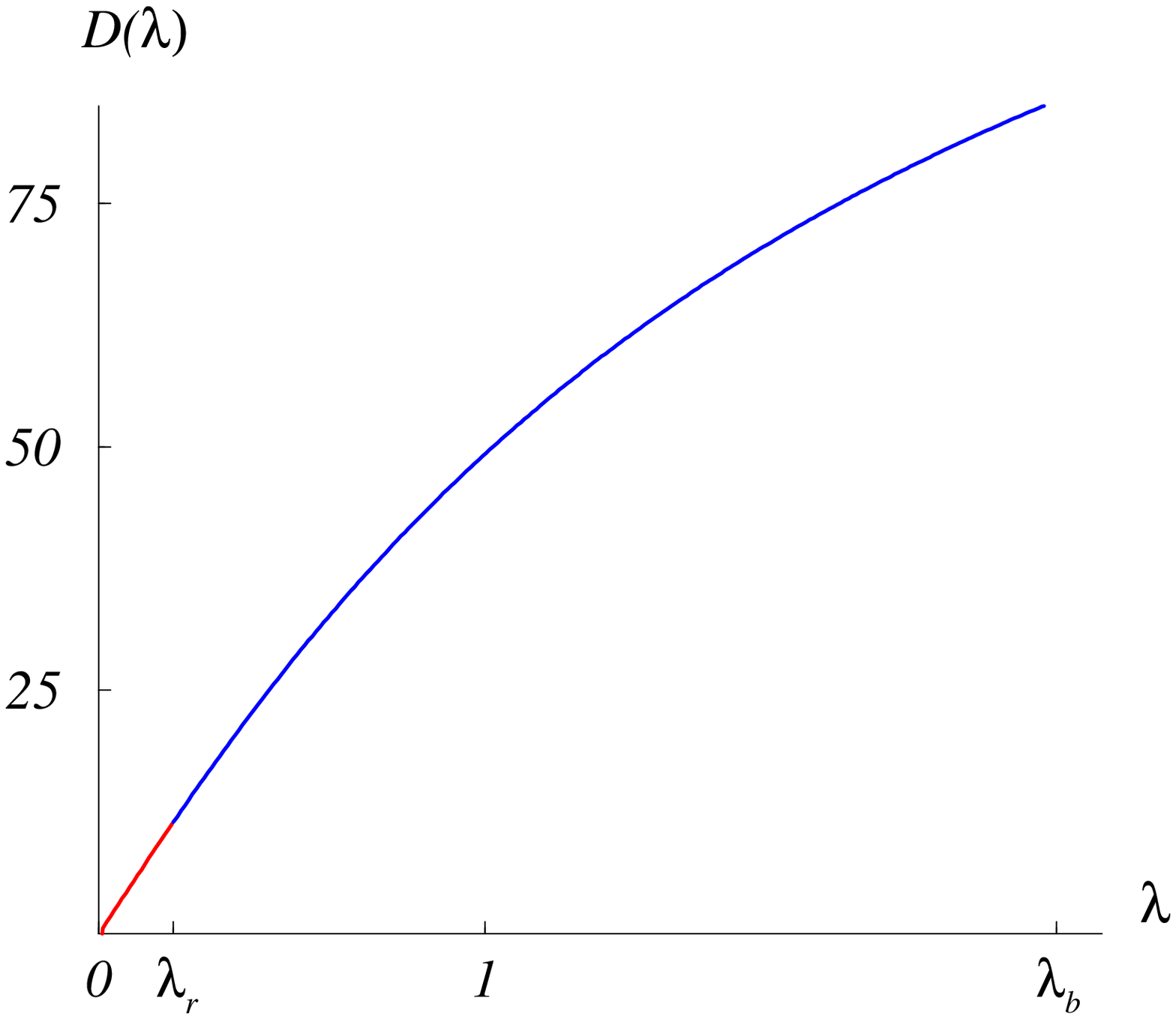, height=2.5in}
\caption{\small Plot of $D(\lambda)$ for dimple-pulse when $v_2(x)$ is odd.   
$a=2.4$, $A=2.8$, $\alpha=0.22$, $u_{\scriptscriptstyle T}=0.18$, $x_{\scriptscriptstyle T}=2.048246$, $\lambda_r=0.192861$, $\lambda_b^{\bf s}=2.48147$. There is no positive  $\lambda$ such that $D(\lambda)=0$, $\lambda \leq \lambda_b^{\bf d}.$ When $v_2(x)$ is odd, $D(\lambda)$ does identify the zero eigenvalue because $D(\lambda)=0$ at $\lambda=0.$ This is consistent with theorem.}
\label{fig:odddimpledet} 
\end{figure}

When there are only two single-pulses, the large pulse could be a dimple-pulse 
instead of a single-pulse. This dimple pulse has the same stability properties as a large pulse.  The parameter set $a=2.4$, $A=2.8$, $\alpha=0.22$, $u_{\scriptscriptstyle T}=0.18$, and $x_{\scriptscriptstyle T}=2.048246$ corresponds to a dimple pulse.  Carrying out the stability calculation yields $D(\lambda)$ shown in Figs.~\ref{fig:evendimpledet} and \ref{fig:odddimpledet}.   We see that there is no zero crossing and thus the dimple pulse is stable.  This is true for all dimple pulses we tested in this category.

\begin{figure}[!htb] 
\centering
\epsfig{figure=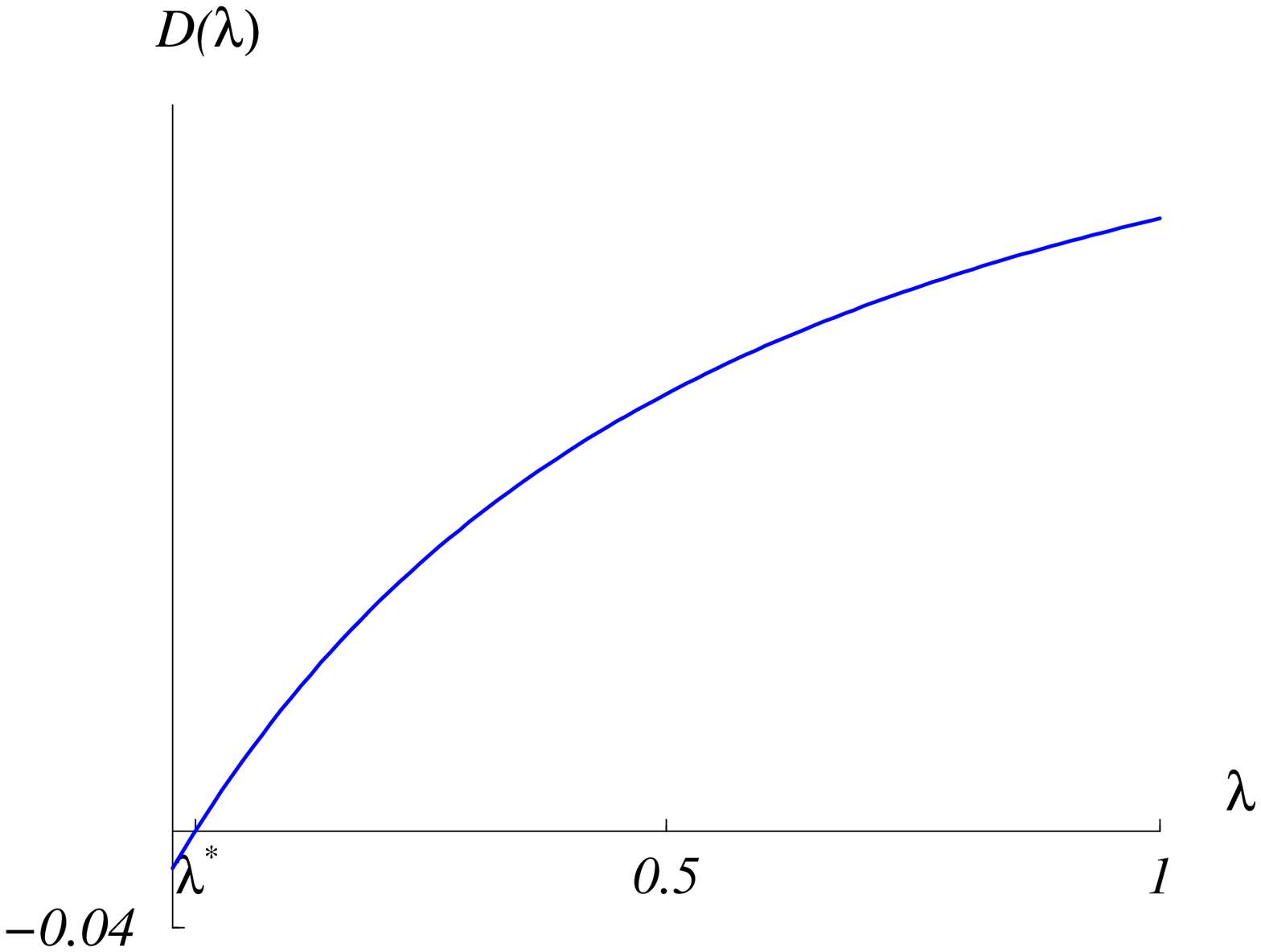, height=2.5in}
\caption{\small Plot of $D(\lambda)$ for the third pulse (a single-pulse) when $v_2(x)$ is even.
$a=2.2$, $A=2.8$, $\alpha=0.8$, $u_{\scriptscriptstyle T}=0.2$, $x_{\scriptscriptstyle T}=2.0629$, 
$c=2.75017$, $D(0)=-0.0153$.  There is a positive $\lambda$ such that $D(\lambda)=0$.}
\label{fig:even3rd1bumpbigdet}
\end{figure}

\begin{figure}[!htb] 
\centering
\epsfig{figure=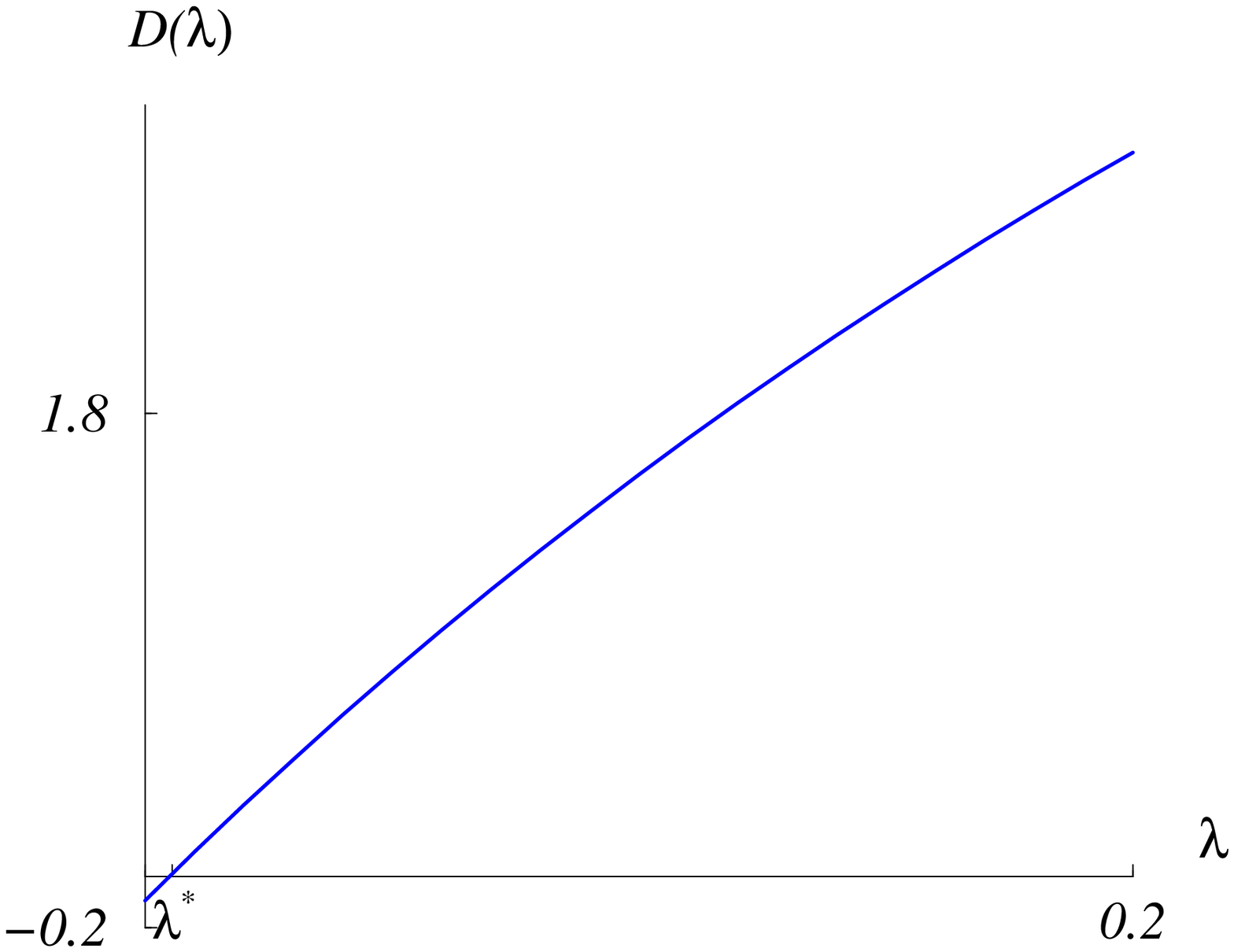, height=2.5in}
\caption{\small Plot of $D(\lambda)$ for the third pulse (a dimple-pulse) when $v_2(x)$ is even.
$a=2.6$, $A=2.8$, $\alpha=0.6178$, $u_{\scriptscriptstyle T}=0.063$, $x_{\scriptscriptstyle T}=1.98232$, 
$c=2.21523$, $D(0)=-0.094$. There is a positive $\lambda$ such that $D(\lambda)=0$.}
\label{fig:even3rddimplebumpdet}
\end{figure}

As shown in \cite{Guo1} and \cite{Guo2}, for certain parameter
regimes, there can be more than two coexisting pulses.  When there are
three pulses, the third pulse can be either a single-pulse or a
dimple-pulse.  
For example, when $A=2.8$, $a=2.2$, $\alpha=0.8$,
$u_{\scriptscriptstyle T}=0.2$, the third pulse is the single-pulse  
\begin{small}
\begin{eqnarray*}
u(x)\hspace{-0.1cm}=\hspace{-0.16cm}\left \{ \begin{array}{ll}
\hspace{-0.25cm} 1.28 \cos (0.47 x) \cosh (1.2x)+1.27\sin(0.47x) \sinh
 (1.2x)+0.8129,  x \in [-x_{\scriptscriptstyle T},
 x_{\scriptscriptstyle T}] 
 \\
\hspace{-0.25cm} 198.78 e^{2|x|}-15.15e^{-|x|},  \mbox {\hspace{6cm} otherwise}
\end{array} \right. \end{eqnarray*}
\end{small}
where $ x_{\scriptscriptstyle T}=2.20629$. $D(\lambda)$ shown in
Fig. \ref{fig:even3rd1bumpbigdet} indicates that this pulse is
unstable. 
When $a=2.6$, $A=2.8$, $\alpha=0.6178$, $u_{\scriptscriptstyle T}=0.063$, the third pulse is the dimple pulse
\begin{small}
\begin{eqnarray*}
\label{eq:3rddimple}
u(x)\hspace{-0.1cm}=\hspace{-0.16cm}\left \{ \begin{array}{ll}
\hspace{-0.25cm} 0.35\cos (1.112 x) \cosh (1.112x)+0.24\sin(1.112x) \sinh (1.112x)+0.163,  x \in [-x_{\scriptscriptstyle T}, x_{\scriptscriptstyle T}]
 \\
\hspace{-0.25cm} 232.89e^{2.6|x|}-9.31e^{-|x|},  \mbox {\hspace{6cm} otherwise}
\end{array} \right. \end{eqnarray*}
\end{small}
where $ x_{\scriptscriptstyle T}=1.98232$.  As seen in Fig.~\ref{fig:even3rddimplebumpdet}, $D(\lambda)$ crosses zero for
 a positive $\lambda$ indicating that it is unstable.  In all the cases that we have examined, we find that the third pulse is unstable.

\section{Double-pulse and its stability}
\label{sec:doublepulse}

\begin{figure}[!htb]
\centering
\includegraphics[height=2.2in, width=4.5in]{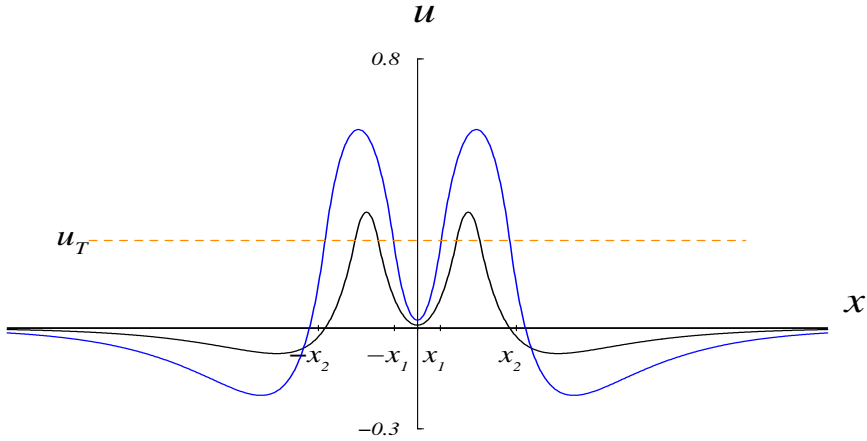}
\caption{\small Double-pulse for Amari case in which
  $\alpha=0$. $A=2.8$, $a=2.6$, $\alpha=0$, $u_{\scriptscriptstyle
    T}=0.26$, $x_1=0.279525$, $x_2=1.20521$.} 
\label{fig:amari2pulse}
\end{figure}

For certain parameter regimes, there can be
double-pulse solutions which have two
disjoint open and finite intervals for which the synaptic input $u(x)$
is above threshold~\cite{Guo1, Guo2, Troy2}.  
An example is shown in 
Fig.~\ref{fig:amari2pulse}).  We consider symmetric double-pulses that
satisfy the  equation 
\begin{eqnarray}
\label{eq:integral_equation2}
u(x)=\int_{-x_2}^{x_1}w(x-y)f[u(y)]dy +\int_{x_1}^{x_2}w(x-y)f[u(y)]dy
\end{eqnarray}
where $x_{1,2}>0$.
Thus 
$u>u_{\scriptscriptstyle T}$ for  $x\in(x_1,x_2) \cup (-x_2, -x_1)$,  
$u=u_{\scriptscriptstyle T}$ for $x=-x_2,-x_1, x_1, x_2$, and
$u<u_{\scriptscriptstyle T}$ outside of these regions and approaches
zero as $x\rightarrow \infty$. 
We  show their existence in \cite{Guo1} and \cite{Guo2}. 

 Linearizing the dynamical neural field equation 
(\ref{eq:int}) around a stationary double-pulse solution $u(x)$ gives 
eigenvalue equation:
\begin{eqnarray}
\label{eq:2bumpeigeneqn1}
(1+\lambda)v(x)  =  w(x-x_1)\frac{v(x_1)}{c_1}+ w(x+x_1)\frac{v(-x_1)}{c_1}
+ w(x-x_2)\frac{v(x_2)}{c_2}\\
 \mbox{} + w(x+x_2)\frac{v(-x_2)}{c_2} +\alpha
 \left(\int_{-x_2}^{-x_1}w(x-y)v(y)dy + 
\int_{x_1}^{x_2}w(x-y)v(y)dy \right) \nonumber
\end{eqnarray}
The eigenvalues $\lambda$ of (\ref{eq:2bumpeigeneqn1}) possess the same
properties as those of the eigenvalue equation for the single-pulse
solutions.   

For simplicity, we consider the Amari case in which $\alpha=0$.  The solution of   (\ref{eq:2bumpeigeneqn1}) for $\alpha>0$ would involve a long calculation.  For $\alpha=0$,  the eigenvalue equation
(\ref{eq:2bumpeigeneqn1}) becomes 
\begin{eqnarray}
\label{eq:amari2bumpeigeneqn1}
(1+\lambda)v(x) = w(x-x_1)\frac{v(x_1)}{c_1}&+ &w(x+x_1)\frac{v(-x_1)}{c_1} \\
& & w(x-x_2)\frac{v(x_2)}{c_2}+w(x+x_2)\frac{v(-x_2)}{c_2} \nonumber
\end{eqnarray}
where $c_1=u'(x_1)$ and $c_2=u'(-x_2)$. Then $u'(-x_1)=-c_1$ and
$u'(x_2)=-c_2.$ 
Using an approach similar to Theorem \ref{the:the1} in the Appendix, we can show 
that $\lambda$ is real. By taking the derivative of (\ref{eq:integral_equation2}), we can also show that zero is an eigenvalue of  system 
(\ref{eq:amari2bumpeigeneqn1}), and the corresponding eigenfunction is
$u'(x).$ 

\begin{figure}[!htb]
\centering
\includegraphics[height=2.4in, width=2.8in]{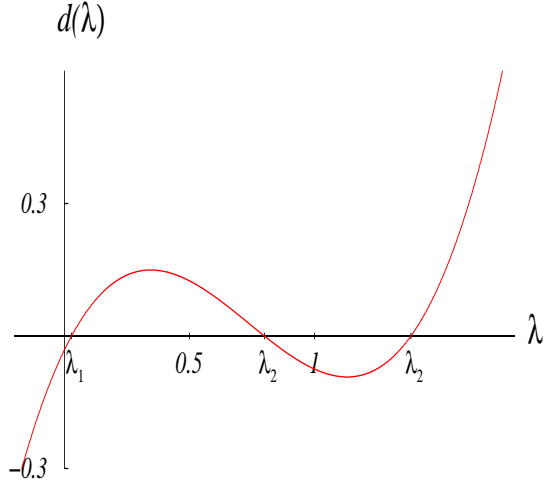}
\caption{\small Plot of polynomial $d(\lambda)$ for the small double-pulse shown in Fig. \ref{fig:amari2pulse}.}
\label{fig:dlambdasmal}
\end{figure}

\begin{figure}[!htb]
\centering
\includegraphics[height=2.4in, width=2.8in]{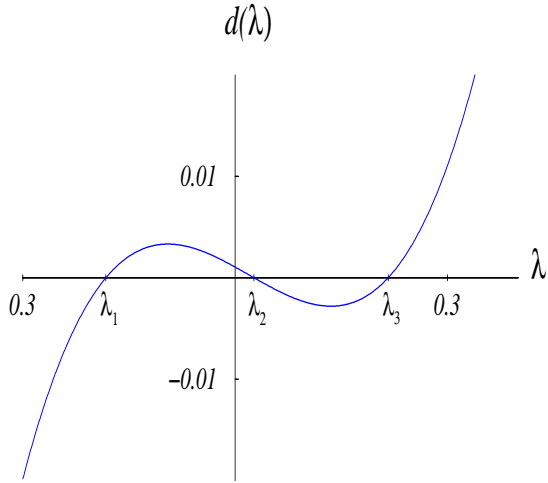}
\caption{\small Plot of polynomial $d(\lambda)$ for the large double-pulse shown in Fig 
\ref{fig:amari2pulse}.}
\label{fig:dlambdabig}
\end{figure}

Setting $x=x_1,$ $x=-x_1,$ $x=x_2$ and $x=-x_2$ in (\ref{eq:amari2bumpeigeneqn1}) gives
a 4-dimensional 
system
\begin{small}
\begin{eqnarray}
\label{eq:2-pulse}
\begin{array}{ll}
\hspace{0.5cm}\left( 
\begin{array}{cccc}
\vspace{0.6cm}
{\displaystyle \frac{w(0)-\lambda-1}{c_1}} & {\displaystyle
  \frac{w(2x_1)}{c_1}} & {\displaystyle \frac{w(x_1-x_2)}{c_2}} &
{\displaystyle \frac{w(x_1+x_2)}{c_2}}\\ 
\vspace{0.6cm}
{\displaystyle \frac{w(2x_1)}{c_1}} & {\displaystyle
  \frac{w(0)-1-\lambda}{c_1}} & {\displaystyle \frac{w(x_1+x_2)}{c_2}}
& {\displaystyle \frac{w(x_1-x_2)}{c_2}}\\ 
\vspace{0.6cm}
{\displaystyle \frac{w(x_1-x_2)}{c_1}} & {\displaystyle
  \frac{w(x_1+x_2)}{c_1}} & {\displaystyle \frac{w(0)-1-\lambda}{c_2}}
& {\displaystyle \frac{w(2x_2)}{c_2}}\\ 
{\displaystyle \frac{w(x_1+x_2)}{c_1}} & {\displaystyle
  \frac{w(x_1-x_2)}{c_1}} & {\displaystyle \frac{w(2x_2)}{c_2}} &
{\displaystyle \frac{w(0)-1-\lambda}{c_2}} 
\end{array}
\right) & 
\hspace{-0.45cm}\left(
\begin{array}{l}
\vspace{0.6cm}
{\displaystyle v(x_1)} \\
\vspace{0.6cm}
{\displaystyle v(-x_1)} \\
\vspace{0.6cm}
{\displaystyle v(x_2)} \\
{\displaystyle v(-x_2)}
\end{array}
\right)
\end{array} \hspace{-0.3cm}=0
\end{eqnarray}
\end{small}
The determinant $D(\lambda)$ of coefficient matrix in system
(\ref{eq:2-pulse}) 
is a fourth order polynomial. Since zero is an eigenvalue, then 
$D(\lambda)= \lambda d(\lambda),$ where $d(\lambda)$ is a third order 
polynomial. Consequently, the stability of the stationary solution $u(x)$ is 
determined by the roots of a third order polynomial
$d(\lambda)$,  which  can be found numerically.  We computed $d(\lambda)$ for the two double pulses shown in Fig.~\ref{fig:amari2pulse}. 
Figure~\ref{fig:dlambdasmal} shows a plot of the third order polynomial $d(\lambda)$ for the small double-pulse. It has three positive zeros indicating instability.  The plot of
$d(\lambda)$ for the
large double-pulse as shown in Fig.~\ref{fig:dlambdabig}) has two positive zeros. Therefore, 
both the small and large double-pulses are
unstable. We have not found any stable double-pulses for any parameter sets that we tested.  However, we have not fully investigated the parameter space of
$A,$ $a$ and $u_{\scriptscriptstyle T}$. 

\section{Discussion}
\label{sec:discussion}
Our results show that although many types of pulse-solutions are
possible, only the family of large pulses and associated dimple pulses
are stable.  For the situation of three coexisting pulses, the third
and largest pulse is always unstable.  It is possible that more than
three pulses can coexist although we did not investigate situations
beyond three.  The double pulses we examined were not stable in
accordance with previous work~\cite{Troy2}.    

The caveat is that we were only able to examine specific examples
individually or over limited parameter ranges.  Although we have an
analytical expression for the eigenvalues  the length of these
expressions makes them difficult to analyze.  As a result, we were
unable to make as strong a claim as Amari who showed that large pulses
are always stable and small pulses are always unstable~\cite{amari77}.
It may be possible to find some patterns in the expressions to make
more general deductions.  From our parameter explorations, we were
unable to find stable pulse solutions other than the large and
associated dimple pulse.    

We wish to note that numerical simulations on discretized lattices can
give misleading results regarding the stability and existence of pulse
solutions of the associated continuum neural field equation.  We
conducted some numerical experiments using a discretization of the
neural field equation (\ref{eq:int}) and to our surprise we were able
to easily find examples of stable dimple and double pulses even though
the continuum analogue shows that these solutions 
either do not exist or cannot be stable.  The resolution to this
paradox is that a discrete lattice may stabilize solutions that are
marginally stable in the continuum case. 

Consider the Amari neural network equation consisting of $N$ neurons
\begin{equation}
\frac{du_i}{dt}=\Delta x \sum_{j=0}^N w(\Delta x(i-j))
\Theta[u_j-u_{\scriptscriptstyle T}], 
\label{eq:nn}
\end{equation}
where $w(i-j)$ is given by  (\ref{eq:coupling}), $\Theta(\cdot)$ is
the Heaviside function, and $\Delta x$ gives the discretization mesh
size.  For an initial condition for which $u_j>u_{\scriptscriptstyle
  T}$ on a contiguous set of points $\{i...k\}$ and $k-i$ is less than
the expected width of the large pulse in the analogous continuum
neural field equation, the numerical solution converges towards the
expected large pulse solution.  However, if the initial set of points
is larger than the width of the large pulse (we have not fully
investigated how much larger it needs to be), then there is a
possibility that the simulation will converge towards an entirely
different state.  

\begin{figure}
\begin{center}
\includegraphics[height=2.2in]{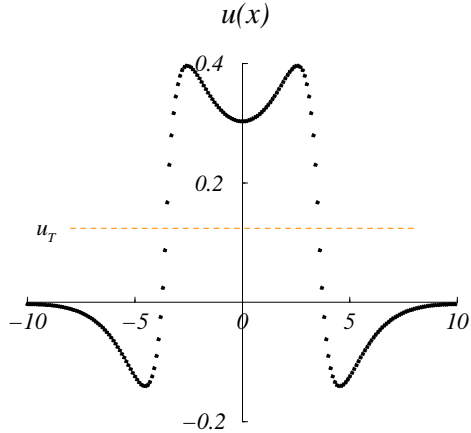}
\end{center}
\caption{Result of numerical simulation of (\ref{eq:nn}) for
  parameters $N=200$, $\Delta x=0.1$, $A=1.8$, $a=1.6$, and
  $u_{\scriptscriptstyle T}=0.124$.  The arbitrary discretization
  length scale is chosen so that $x=0.1i$.}  
\label{fig:margindimple}
\end{figure}

For example, a numerical simulation of the parameter set  $N=200$,
$\Delta x=0.1$, $A=1.8$, $a=1.6$, and $u_{\scriptscriptstyle T}=0.124$
with an initial condition $u_i=1$ for $i\in{50...150}$, converges to a
stable dimple-pulse state shown in Fig.~\ref{fig:margindimple}.
Different initial domains will lead to different attracting states
where the width is close to the initial domain width.  For a large
enough initial domain,  the dimple pulse will break into a stable
double-pulse.  Increasing the initial domain can lead to increasingly
higher number stable multiple pulses. 

\begin{figure}
\begin{center}
\includegraphics[height=2.4in]{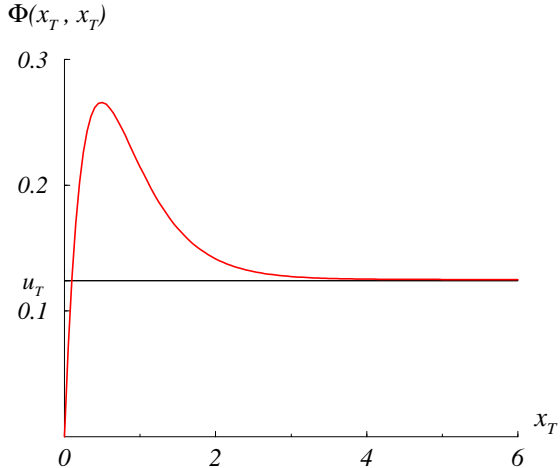}
\end{center}
\caption{Existence condition for pulse solutions of neural field
  equation (\ref{eq:nnfield}) for parameters  
$A=1.8$, $a=1.6$, and $u_{\scriptscriptstyle T}=0.124$. }
\label{fig:amariphi}
\end{figure}

We can show that these states do not exist in the analogous continuum
neural field equation 
Consider a stationary pulse solution of (\ref{eq:int}) for $\alpha=0$.
A pulse of width $x_{\scriptscriptstyle T}$ satisfies  
\begin{equation}
u(x)=\phi(x,x_{\scriptscriptstyle T}),
\label{eq:nnfield}
\end{equation}
where
\begin{equation}
\phi(x,x_{\scriptscriptstyle T})=\int_{-x_{\scriptscriptstyle
    T}}^{x_{\scriptscriptstyle T}} A e^{-a|x-y|} - e^{-|x-y|} dy. 
\end{equation}
The pulse can exist if it satisfies the existence condition
\begin{equation}
 u_{\scriptscriptstyle T}=\phi(x_{\scriptscriptstyle
 T},x_{\scriptscriptstyle T}) 
 \end{equation}
{}from which the width $x_{\scriptscriptstyle T}$ can be obtained.
A plot of the existence condition is shown in Fig.~\ref{fig:amariphi}.

It is immediately apparent that the large pulse does not exist.  The
existence function approaches $u=u_{\scriptscriptstyle T}$ from above
for large enough $x_{\scriptscriptstyle T}$.  While it is very close
to $u_{\scriptscriptstyle T}$ it never crosses it.   However, for the
analogous discretized equation (\ref{eq:nn}), the discrete mesh 
can break the symmetry of
this nearly marginal mode and result in a family of stable pulse
solutions for arbitrary widths larger than a given width.  

This effect can be intuitively understood by examining
Fig.~\ref{fig:margindimple}.  The neurons immediately adjacent to the
edge of the pulse are significantly below threshold and thus have no
effect on the rest of the network.  A perturbation on the order of the
distance they are below threshold would be necessary to cause these
neurons to fire and
influence the network.  In the continuum equation, the neurons on the
boundary of the pulse are precisely at threshold.  Arbitrarily small
perturbations can push the field above threshold and influence the
other neurons.  A stable pulse must withstand these edge
perturbations.  Discretization eliminates these destabilizing edge
perturbation effects. 

We can make a simple estimate of how fine the discretization mesh must
be in order for these discrete affects to disappear.  The distance the
neuron adjacent to the pulse is below threshold is approximately given
by $\partial_x \phi(x=x_{\scriptscriptstyle T},x_{\scriptscriptstyle
  T}) dx\sim (A-1) dx$. 
For the parameter set of our simulation, the continuum existence
condition shows that $\phi(x_{\scriptscriptstyle
  T},-x_{\scriptscriptstyle T})-u_{\scriptscriptstyle T}> 0.001$.
Thus  to eliminate the discreteness effect, we require the adjacent
neuron to be above threshold. i.e. $(A-1)dx<0.001$ as it would be in
the continuum case.  This leads to an estimate of $dx<0.00125$.
Hence, for a domain of dimension $x>20$, a network size of $N>16,000$
is necessary to eliminate the discreteness effect. 

Biological neural networks are inherently discrete.  Thus this
discreteness effect may be exploited by the brain to stabilize
localized excitations.  Our numerical simulation is an example of a
discretized line attractor~\cite{seung} where the width of the pulse
is determined by the initial condition.  Although, the discrete
network may have a richer structure, this does not imply that the
study of continuum neural field equations are not necessary.  Field
equations lend themselves more readily to analysis and many insights
into the structure and properties of neural networks have been gained
by studying them.  We suggest that studies combining neural field
equations, discrete neural network equations and biophysically based
spiking neurons may be a fruitful way to uncover the dynamics of these
systems.

\section{Appendix 1: Properties of the eigenvalue problem}
\label{sec:property}

We prove some properties of the 
the eigenvalue problem (\ref{eq:eigenvalueequation}) with the
connection function given by (\ref{eq:coupling}). 
First consider functions
$$\phi_1(x)=\frac{1}{2a}\int_{-\infty}^{\infty}e^{-a|x-y|}
(F_u+\Theta_u)v(y)dy$$   
$$\phi_2(x)=\frac{1}{2}\int_{-\infty}^{\infty}e^{-|x-y|}(F_u+\Theta_u)v(y)dy$$
where $F(u)=\alpha(u-u_{\scriptscriptstyle T})$, $\Theta(u)$ is the 
Heaviside function, and subscript denotes partial differentiation.

\begin{lemma}
\label{lem:lem3}
The eigenfunction $v(x)$ satisfies
$$(1+\lambda)v=2(aA\phi_1-\phi_2)$$
\end{lemma}
\begin{proof}
\begin{eqnarray*}
(1+\lambda)v & = & w(x-x_{\scriptscriptstyle
T})\frac{v(x_{\scriptscriptstyle T})}{c}+w(x+x_{\scriptscriptstyle
T})\frac{v(-x_{\scriptscriptstyle T})}{c}+\alpha
\int_{-x_{\scriptscriptstyle T}}^{x_{\scriptscriptstyle
T}}w(x-y)v(y)dy \\
& = & \int_{-\infty}^{\infty}
w(x-y)\frac{\delta(x-x_{\scriptscriptstyle
T})+\delta(x+x_{\scriptscriptstyle
T})}{c}v(y)dy+\int_{-\infty}^{\infty}w(x-y)F_u v(y)dy \\
& = &\int_{-\infty}^{\infty}
w(x-y)\Theta_u v(y)dy+\int_{-\infty}^{\infty}w(x-y)F_u v(y)dy \\
& = &A
\int_{-\infty}^{\infty}e^{-a|x-y|}(F_u+\Theta_u)v(y)dy -
\int_{-\infty}^{\infty}e^{-|x-y|}(F_u+\Theta_u)v(y)dy   
\\
& = & 2(aA\phi_1-\phi_2) 
\end{eqnarray*} 
\end{proof}

\begin{lemma}
\label{lem:lem4}
Functions $\phi_1$ and $\phi_2$ satisfy
\begin{eqnarray}
\label{eq:localeq4}
-\phi_1''+a^2\phi_1 & = & (F_u+\Theta_u)v \\
\label{eq:localeq5}
-\phi_2''+a^2\phi_2 & = & (F_u+\Theta_u)v
\end{eqnarray}
\end{lemma}
\begin{proof}
The second order derivative of $\phi_1(x)$ is 
\begin{eqnarray}
\label{eq:localeq2}
 \label{eq:localeq3} 
\hspace{1cm}
 \phi_1''& = & \frac{a}{2}\left[\int_{-\infty}^{x}e^{-a(x-y)}(F_u+\Theta_u)v
dy+\int_{x}^{\infty}e^{a(x-y)}(F_u+\Theta_u)v dy  \right]\\
& & \mbox{} -(F_u+\Theta_u)v \nonumber
\end{eqnarray}
$-(\ref{eq:localeq3})+a^2 \phi_1(x)$ yields
\begin{equation}
-\phi_1''+a^2\phi_1  =  (F_u+\Theta_u)v.
\end{equation}
$-\phi_2''+a^2\phi_2 =  (F_u+\Theta_u)v$ can be obtained in the same
fashion.
\end{proof}
\begin{lemma}
\label{lem:lem5} ${\displaystyle \lim_{x \rightarrow \pm \infty}
  \phi_{1,2} = 0}$ 
and ${\displaystyle \lim_{x \rightarrow \pm \infty} \phi'_{1,2} = 0}$ provided
that $v(x)$ is bounded on $(-\infty, \infty)$ and exponentially
decays to zero as $x \rightarrow \pm \infty$
\end{lemma} 
\begin{proof}
 When $x>>x_{\scriptscriptstyle T}$
\begin{eqnarray*}
\phi_1(x) 
& = &\frac{1}{2a}\left[\alpha e^{-ax}\int_{-x_{\scriptscriptstyle
T}}^{x_{\scriptscriptstyle
T}}e^{ay}v(y)dy+e^{-a(x-x_{\scriptscriptstyle
T})}\frac{v(x_{\scriptscriptstyle T})}{c}+e^{-a(x+x_{\scriptscriptstyle
T})}\frac{v(-x_{\scriptscriptstyle T})}{c}\right].
\end{eqnarray*}
Hence, ${\displaystyle \lim_{x \rightarrow \infty} \phi_1=0}$ 
provided that $v(x)$ is bounded on $\left[-x_{\scriptscriptstyle
T},x_{\scriptscriptstyle T}\right]$.

When $x<<-x_{\scriptscriptstyle T}<0$, as $x \rightarrow -\infty$
\begin{eqnarray*}
\phi_1(x)=\frac{1}{2a}\left[\alpha e^{ax}\int_{-x_{\scriptscriptstyle
T}}^{x_{\scriptscriptstyle
T}}e^{-ay}v(y)dy+e^{a(x-x_{\scriptscriptstyle
T})}\frac{v(x_{\scriptscriptstyle T})}{c}+e^{a(x+x_{\scriptscriptstyle
T})}\frac{v(-x_{\scriptscriptstyle T})}{c}\right]\rightarrow 0 
\end{eqnarray*}

\begin{eqnarray*}
\phi'_1=\frac{1}{2}\left[-\int_{-\infty}^{x}e^{-a(x-y)}(F_u+\Theta_u)v
dy+\int_{x}^{\infty}e^{a(x-y)}(F_u+\Theta_u)v dy\right] 
\end{eqnarray*}
As $x \rightarrow \infty$,
\begin{eqnarray*}
\lim_{x \rightarrow \infty} \phi_1' & = & \lim_{x \rightarrow
\infty}\left\{-\frac{1}{2}\int_{-\infty}^{x}e^{-a(x-y)}(F_u+\Theta_u)v
dy \right\} \\ 
& = & \lim_{x \rightarrow \infty}\left\{-\frac{e^{-ax}}{2}
\left[\alpha \int_{-x_{\scriptscriptstyle T}}^{x_{\scriptscriptstyle T}}
e^{ay}dy + e^{ay}\frac{v(x_{\scriptscriptstyle
T})}{c}\right]\right\}=0
\end{eqnarray*}
As $x \rightarrow -\infty$
\begin{eqnarray*}
\lim_{x \rightarrow -\infty} \phi_1' & = & \lim_{x \rightarrow
-\infty}\left\{-\frac{1}{2}\int_{x}^{\infty}e^{a(x-y)}(F_u+\Theta_u)v
dy \right\} \\ 
& = & \lim_{x \rightarrow \infty}\left\{-\frac{e^{ax}}{2}
\left[\alpha \int_{-x_{\scriptscriptstyle T}}^{x_{\scriptscriptstyle T}}
e^{ay}dy + e^{ax_{\scriptscriptstyle T}}\frac{v(x_{\scriptscriptstyle
T})}{c}\right]\right\}=0
\end{eqnarray*}
Similarly, one can prove that ${\displaystyle \lim_{x \rightarrow \pm
\infty}\phi_{2}=0}$ and ${\displaystyle \lim_{x \rightarrow \pm
\infty}\phi_{2}'=0}$.
Therefore, 

${\displaystyle \lim_{x \rightarrow \pm
    \infty}\phi_{1,2}=0}$ and ${\displaystyle \lim_{x \rightarrow \pm\infty} \phi'_{1,2}=0}$.
\end{proof}
  
\begin{theorem}
\label{the:the1} The eigenvalue $\lambda$ in (\ref{eq:eigenvalueequation}) is
always real.
\end{theorem}
\begin{proof} Using the results of Lemma \ref{lem:lem4},
$aA\bar{\phi_1}(\ref{eq:localeq4})-\bar{\phi_2}(\ref{eq:localeq5})$ gives
\begin{eqnarray}
\label{eq:local6} aA\bar{\phi_1}(-\phi_1''+a^2
\phi_1)-\bar{\phi_2}(-\phi_2''+\phi_2) & = &
(F_u+\Theta_u)v(aA\bar{\phi_1}-\bar{\phi_2})
\end{eqnarray}
where $\bar{\phi}_{1,2}$ are the complex conjugates of
$\phi_{1,2}.$ Integration by parts gives
\begin{eqnarray*}
\int_{-\infty}^{\infty}\bar{\phi_1}\phi_1''dx  = 
\bar{\phi_1}\phi_1'|_{-\infty}^{\infty}-\int_{-\infty}^{\infty}
\bar{\phi_1'}\phi_1'dx=
-\int_{-\infty}^{\infty}\left|\phi_1'\right|^{2}dx 
\end{eqnarray*}
and similarly 
${\displaystyle \int_{-\infty}^{\infty}\bar{\phi_2}\phi_2''dx  =-
\int_{-\infty}^{\infty}\left|\phi_2'\right|^{2}dx}.$
>From Lemma \ref{lem:lem3}
\begin{eqnarray*}
\frac{1}{2}(1+ \lambda)v & = & aA\phi_1-\phi_2 \\
\frac{1}{2}(1+\bar{\lambda})\bar{v} & = &
aA\bar{\phi_1}-\bar{\phi_2}
\end{eqnarray*}
Integrating both sides of (\ref{eq:local6}) gives
\begin{small}
 \begin{eqnarray}
\lefteqn{aA\left(\int_{-\infty}^{\infty} \left|\phi_1'
\right|^2 dx + a^2\int_{-\infty}^{\infty} \left|\phi_1 \right|^2
dx \right)-} \\
&& \left(\int_{-\infty}^{\infty} \left|\phi_2' \right|^2
dx + \int_{-\infty}^{\infty} \left|\phi_2 \right|^2 dx \right) =
\frac{1}{2}(1+\bar{\lambda})\int_{-\infty}^{\infty}\left|v\right|^2
(F_u+\Theta_u)dx   \nonumber
\label{eq:normeq}
\end{eqnarray}
\end{small}
Using 
$${\displaystyle \int_{-\infty}^{\infty}\left|v\right|^2\Theta_udx =
  \frac{1}{c}\int_{-\infty}^{\infty}\left|v\right|^2
  \left(\delta(x-x_{\scriptscriptstyle   
T})+\delta(x+x_{\scriptscriptstyle T})\right)dx=\frac{1}{c}\left( 
\left|v(x_{\scriptscriptstyle 
T})\right|^2+\left|v(-x_{\scriptscriptstyle T})\right|^2\right)}$$
in (\ref{eq:normeq}) and rearranging gives
\begin{small}
\begin{eqnarray}
\label{eq:eigenvalue}
\hspace{0.5cm}
\frac{1}{2}(1+\bar{\lambda})=\frac{aA\left(\int_{-\infty}^{\infty}
\left|\phi_1' \right|^2 dx + a^2\int_{-\infty}^{\infty}
\left|\phi_1 \right|^2 dx \right)-\left(\int_{-\infty}^{\infty}
\left|\phi_2' \right|^2 dx + \int_{-\infty}^{\infty} \left|\phi_2
\right|^2 dx
\right)}{\int_{-\infty}^{\infty}F_u\left|v\right|^2 dx+\frac{1}{c}\left(
\left|v(x_{\scriptscriptstyle
T})\right|^2+\left|v(-x_{\scriptscriptstyle T})\right|^2\right)}
\end{eqnarray}
\end{small}
The right-hand side of (\ref{eq:eigenvalue}) is real, therefore
$\lambda$  is real. 
\end{proof}

\begin{theorem}
\label{the:the2} The eigenvalue $\lambda$ in (\ref{eq:eigenvalueequation}) is
bounded by $\lambda_b \equiv \frac{2k_0}{c}+ 2 \alpha k_1
x_{\scriptscriptstyle 
T}-1$ where $k_0$ is the maximum of $\left|w(x)\right|$ on $\left[0,
2x_{\scriptscriptstyle T}\right]$ and $\left|w(x-y)\right|\leq k_1$ for
all $(x,y) \in J \times J,$ where $J=\left[-x_{\scriptscriptstyle
    T},x_{\scriptscriptstyle T}\right].$ 
\end{theorem}

\begin{proof}
We write the eigenvalue problem 
(\ref{eq:eigenvalueequation}) as
\begin{eqnarray}
\label{eq:localeq7} (1+\lambda)v=Lv
\end{eqnarray}
where operator $L$ is defined as (\ref{eq:operatorL}).

Function $w(x-y)$ is continuous on square $J
\times J.$ 
We take the norm of both sides of (\ref{eq:localeq7})
\begin{eqnarray*}
(1+\lambda)\|v\| = \|L v\|
\end{eqnarray*}
with norm 
\begin{eqnarray*}
\|\cdot\|=\max_{x \in J}| \cdot |
\end{eqnarray*}
Thus
\begin{eqnarray*}
\|L v\| & = &\left \|w(x-x_{\scriptscriptstyle
T})\frac{v(x_{\scriptscriptstyle T})}{c}+w(x+x_{\scriptscriptstyle
T})\frac{v(-x_{\scriptscriptstyle T})}{c}+\alpha
\int_{-x_{\scriptscriptstyle T}}^{x_{\scriptscriptstyle
T}}w(x-y)v(y)dy\right\| \\
& \leq & \max_{x \in J}\left|w(x-x_{\scriptscriptstyle
T})\frac{v(x_{\scriptscriptstyle T})}{c}\right|+\max_{x \in
J}\left|w(x+x_{\scriptscriptstyle
T})\frac{v(-x_{\scriptscriptstyle T})}{c}\right| + \\
&& \hspace{1cm}\max_{x \in J}\left|\alpha \int_{-x_{\scriptscriptstyle
T}}^{x_{\scriptscriptstyle T}}w(x-y)v(y)dy\right| \\
& \leq &\left|w(x-x_{\scriptscriptstyle T})\right|
\frac{\|v\|}{c}+\left|w(x+x_{\scriptscriptstyle
T})\right|\frac{\|v\|}{c} + \alpha 
\left\| v\right\|\int_{-x_{\scriptscriptstyle T}}^{x_{\scriptscriptstyle
T}}\max_{x \in J}\left|w(x-y)\right|dy \\
& \leq &2k_0 \frac{\|v(x)\|}{c}+ 2 \alpha k_1
x_{\scriptscriptstyle T}\|v(x)\|
\end{eqnarray*}
where $$k_0=\max_{x \in J}\left|w(x-x_{\scriptscriptstyle
T})\right|=\max_{x \in J}\left|w(x+x_{\scriptscriptstyle
T})\right|$$ since $w(x)$ is symmetric and
$\left|w(x-y)\right|\leq k_1$ for all $(x,y) \in J \times J.$
Therefore
$$
(1+ \lambda)\|v(x)\|=\|L v(x)\| \leq  2k_0 \frac{\|v(x)\|}{c}+
2
\alpha k_1 x_{\scriptscriptstyle T}\|v(x)\|
$$
leading to
$$
\lambda \leq  \frac{2k_0}{c}+ 2 \alpha k_1
x_{\scriptscriptstyle T}-1\equiv \lambda_b. {\mbox{\qquad}}
$$
\end{proof}

\begin{theorem}
\label{the:the9}
 $\lambda=0$ is an eigenvalue.
\end{theorem}
\begin{proof}
Consider the equilibrium equation 
\begin{eqnarray}
\label{eq:equilibriumeqn}
u(x) &  = & \int_{-\infty}^{\infty}w(x-y)f[u(y)]dy \nonumber \\
& = & \int_{-x_{\scriptscriptstyle T}}^{x_{\scriptscriptstyle T}}w(x-y) \left\{\alpha \left[u(y)-u_{\scriptscriptstyle T}\right]+1 \right\}dy 
\end{eqnarray}
where $u(x)$ is a stationary standing pulse solution.
After a change of variables $p=x-y$,  (\ref{eq:equilibriumeqn}) becomes
\begin{eqnarray}
\label{eq:changeofvariable}
u(x)=\int_{x-x_{\scriptscriptstyle T}}^{x+x_{\scriptscriptstyle T}}w(p) 
\left\{\alpha \left[u(x-p)-u_{\scriptscriptstyle T}\right]+1 \right\}dp
\end{eqnarray}
Differentiating (\ref{eq:changeofvariable}) with respect to $x$ yields
\begin{eqnarray}
\label{eq:eigenprob}
\hspace{0.6cm} u'(x) & = & w(x+x_{\scriptscriptstyle T})\left[\alpha
  (u(-x_{\scriptscriptstyle T})- 
u_{\scriptscriptstyle T})+1\right]- w(x-x_{\scriptscriptstyle T})\left[\alpha 
(u(x_{\scriptscriptstyle T})-u_{\scriptscriptstyle T})+1\right] \\ &
  & \mbox{} + \alpha \int_{x-x_{\scriptscriptstyle
  T}}^{x+x_{\scriptscriptstyle T}}  
w(p)u'(x-p)dp\nonumber  
\end{eqnarray}
Since $u(-x_{\scriptscriptstyle T})=u(x_{\scriptscriptstyle T})u_{\scriptscriptstyle T}$ and $u'(-x_{\scriptscriptstyle T})=c=-u'(x_{\scriptscriptstyle T}),$ 
\begin{eqnarray}
u'(x) & = & w(x+x_{\scriptscriptstyle T})\frac{u'(-x_{\scriptscriptstyle T})}{c}- w(x-x_{\scriptscriptstyle T}) \frac{-u'(x_{\scriptscriptstyle T})}{c}+
\alpha \int_{x-x_{\scriptscriptstyle T}}^{x+x_{\scriptscriptstyle T}} 
w(p)u'(x-p)dp  \nonumber \\
& = & w(x-x_{\scriptscriptstyle T})\frac{u'(x_{\scriptscriptstyle T})}{c}+w(x+x_{\scriptscriptstyle T})\frac{u'(-x_{\scriptscriptstyle T})}{c}+
\alpha \int_{-x_{\scriptscriptstyle T}}^{x_{\scriptscriptstyle T}} 
w(x-y)u'(y)dy 
\label{eq:eigenprob1}
\end{eqnarray}
(\ref{eq:eigenprob1}) is the eigenvalue problem (\ref{eq:eigenvalueequation}) with eigenvalue $\lambda$ satisfying $1+\lambda=1$, resulting in $\lambda=0.$ The corresponding eigenfunction is $u'(x)$. Therefore, $\lambda=0$ is an eigenvalue of (\ref{eq:eigenvalueequation}) corresponding to eigenfunction $u'(x).$  \qquad
\end{proof}

\begin{theorem}
\label{the:theapp} 
Consider the operator
\begin{equation}
L=T_1+T_2,
\end{equation}
where
\begin{eqnarray}
T_1 (v(x)) &=& w(x-x_{\scriptscriptstyle
T})\frac{v(x_{\scriptscriptstyle T})}{c}+w(x+x_{\scriptscriptstyle
T})\frac{v(-x_{\scriptscriptstyle T})}{c}, \mbox{\hspace{0.6cm}}
T_1:C\left[-x_{\scriptscriptstyle T},x_{\scriptscriptstyle T}\right]
\rightarrow C\left[-x_{\scriptscriptstyle T},x_{\scriptscriptstyle
T}\right] \nonumber\\ 
T_2 (v(x)) &=&\alpha
\int_{-x_{\scriptscriptstyle T}}^{x_{\scriptscriptstyle
T}}w(x-y)v(y)dy, \mbox{\hspace{2.8cm}} T_2:C\left[-x_{\scriptscriptstyle
T},x_{\scriptscriptstyle T}\right] \rightarrow
C\left[-x_{\scriptscriptstyle T},x_{\scriptscriptstyle T}\right]
\nonumber
\end{eqnarray}
Both $T_1$ and $T_2$ and hence $L$ are compact operators.

\end{theorem}
\begin{proof}  It is obvious that both $T_1$ and $T_2$ are linear operators. 
The boundedness of $T_1$ follow from
\begin{eqnarray*}
\|T_1v\| & = & \max_{x \in J}\left|w(x-x_{\scriptscriptstyle T})\frac{v(x_{\scriptscriptstyle T})}{c}+w(x+x_{\scriptscriptstyle T})\frac{v(-x_{\scriptscriptstyle T})}{c} \right|\\
& \leq & |w(x-x_{\scriptscriptstyle T})| \frac{\|v(x)\|}{c}+\left|w(x+x_{\scriptscriptstyle T})\right|\frac{\|v(x)\|}{c} \\
& \leq & 2k_0 \frac{\|v\|}{c}
\end{eqnarray*}
Let $v_n$ be any bounded sequence in  $C\left[-x_{\scriptscriptstyle
    T},x_{\scriptscriptstyle T}\right]$ and $\|v_n\| \leq c_0$ for all
$n.$ Let $y_n^1=T_1v_n.$ Then $\|y_n^1\| \leq \|T_1\|\|v_n\|.$ Hence
sequence ${y_n^1}$ is bounded. Since $w(x,t)=w(x-t)$ is continuous on
$J \times J$ and $J \times J$ is compact, $w$ is uniformly continuous
on $J \times J.$ Hence, for any given $\epsilon_1 >0,$ there is a
$\delta_1>0$ such that for $t=x_{\scriptscriptstyle T}$ and all $x_1,
x_2 \in J$ satisfying $\left|x_1-x_2 \right|< \delta_1$
$$\left|w(x_1-x_{\scriptscriptstyle T})-w(x_2-x_{\scriptscriptstyle
  T})\right|<\frac{c}{2 c_0}\epsilon_1.$$ 

Consequently, for $x_1$, $x_2$ as before and every $n,$ one can obtain 
\begin{eqnarray*}
\left|y_n^1(x_1)-y_n^1(x_2)\right| &=& \left|\left[w(x_1-x_{\scriptscriptstyle T})-w(x_2-x_{\scriptscriptstyle T})\right]\frac{v_n(x_{\scriptscriptstyle T})}{c} \right.\\
& + & \left. \left[w(x_1+x_{\scriptscriptstyle T})-w(x_2+x_{\scriptscriptstyle T})\right]\frac{v_n(-x_{\scriptscriptstyle T})}{c}\right| \\
& < & \left|w(x_1-x_{\scriptscriptstyle T})-|w(x_2-x_{\scriptscriptstyle T})\right|\frac{c_0}{c}+\left|w(x_1+x_{\scriptscriptstyle T})-w(x_2+x_{\scriptscriptstyle T})\right|\frac{c_0}{c} \\
 & < & \frac{c}{2 c_0}\epsilon_1 \frac{c_0}{c} + \frac{c}{2 c_0} \epsilon_1\frac{c_0 }{c} =\epsilon_1
\end{eqnarray*}
Boundedness of $T_2$ follows from 
\begin{eqnarray*}
\|T_2 v\| \leq \|v\| \max_{x\in J}\int_{-x_{\scriptscriptstyle T}}^{x_{\scriptscriptstyle T}} |w(x-t)|dt
\end{eqnarray*}
Similarly, let $y_n^2=T_2v_n$. Then $y_n^2$ is bounded. For any given $\epsilon_2 >0,$ there is a $\delta_2>0$ such that 
for any $t \in J$ and all $x_1, x_2 \in J$ satisfying $\left|x_1-x_2 \right|< \delta_2$ $$\left|w(x_1-t)-w(x_2-t)\right|<\frac{\epsilon_2}{2 x_{\scriptscriptstyle T}}$$
\begin{eqnarray*}
\left|y_n^2(x_1)-y_n^2(x_2)\right| &=& \left| \int_{-x_{\scriptscriptstyle T}}^{x_{\scriptscriptstyle T}}[w(x_1-t)-w(x_2-t)]v_n(t)dt\right|\\
 & < & 2x_{\scriptscriptstyle T}\frac{\epsilon_2}{2x_{\scriptscriptstyle T} c_0}=\epsilon_2
\end{eqnarray*}

This proves the equicontinuity of $\{y_n^1\}$ and $\{y_n^2\}.$ By
Ascoli's theorem, both sequences have  convergent
subsequences. ${v_n}$ is an arbitrary bounded sequence and $y_n^1=T_1
v_n,$  $y_n^2=T_2 v_n.$  
The compactness of $T_1$  and $T_2$ follows from the criterion that an
operator is compact if and only if it maps every bounded sequence
${x_n}$ in $X$ onto a sequence ${Tx_n}$ in $Y$ which has a convergent
subsequence.  
\end{proof}

\begin{theorem}
\label{the:the5} $\lambda=-1$ is the only possible accumulation point
of the eigenvalues of $L$ and every spectral value $\lambda \neq -1$
of $L$ is an eigenvalue of $L$.  Thus the only possible essential
spectrum of compact operator $L$ is at $\lambda=-1$. 
\end{theorem}
\begin{proof}
Let $\gamma=(1+\lambda)$, the eigenvalue problem becomes
$$\gamma v(x)=L v(x),$$ and the linear operator $L$ is compact on the normed 
space $C\left[-x_{\scriptscriptstyle
T},x_{\scriptscriptstyle T}\right].$ $\gamma$ is the eigenvalue of operator 
$L$. The spectrum of a compact operator is a countable set with no accumulation
point different from zero. Each nonzero member of the spectrum is an
eigenvalue of the  
compact operator with finite multiplicity \cite{Functional3,Kato}. Therefore,
the only possible point of accumulation for the spectrum set of compact 
operator $L$ is $\gamma=0$, $i.e.,$ $\lambda=-1$  and every spectral value 
$\lambda \neq -1$ of $L$ is an eigenvalue of $L$.
This suggests that the only possible essential spectrum is at
$\lambda=-1$. All the  
spectral values $\lambda$ such that $\lambda>-1$ are eigenvalues.
\end{proof}

\begin{lemma}
\label{lem:lem8} The zero of $B$,
$\lambda_B$, obeys
$-1<\lambda_B<\lambda_r$. For the case $a^3>A$, $\lambda_l<\lambda_B<
\lambda_r$, and for the case $a^3<A$, $\lambda_B<\lambda_l<\lambda_r.$
\end{lemma}
\begin{proof}
Set
$$B=(1+\lambda)(a^2+1)+2\alpha(1-aA)=0$$
The zero of $B$ is
$$\lambda_B=-\frac{a^2+1+2\alpha-2aA\alpha}{a^2+1}=-1+\frac{2\alpha(aA-1)}{a^2+1}>-1.$$
$\Delta$ is a quadratic function in $\lambda$ and it has two
zeros.
The left zero is
$$\lambda_l=\frac{1-a^2+2aA\alpha+2\alpha-4\alpha \sqrt{aA}}{a^2-1}$$
The right zero is
$$\lambda_r=\frac{1-a^2+2aA\alpha+2\alpha+4\alpha
\sqrt{aA}}{a^2-1}$$
The difference between $\lambda_r$ and
$\lambda_B$ is
$$\lambda_r-\lambda_B=\frac{4a \alpha(a+A)+4 \alpha \sqrt{aA}(a^2+1)}{a^4-1}>0$$
Therefore $-1<\lambda_B<\lambda_r$.

The difference between $\lambda_B$ and $\lambda_l$ is
${\displaystyle \lambda_B-\lambda_l=\frac{4\alpha(\sqrt{aA}-1)(a^2-\sqrt{aA})}{a^4-1}}.$
The sign of $\lambda_B-\lambda_l$ depends on $a^2-\sqrt{aA}$.
If $a^2-\sqrt{aA}$ is positive, $i.e.$  $a^3>A$, then
$\lambda_l<\lambda_B<\lambda_r$.
If $a^2-\sqrt{aA}$ is negative, $i.e.$, $a^3<A$, then
$\lambda_B<\lambda_l<\lambda_r.$ 
 \end{proof}
\begin{lemma}
\label{lem:lem9}
(i)  For $a^3>A$ and
$\lambda_l<\lambda_B<\lambda_r$, $B$ does not intersect the left branch or the right branch of $\sqrt{\Delta}$.
(ii) For  $a^3<A$ and
$\lambda_B<\lambda_l<\lambda_r,$ $B$ intersects only the left branch of $\sqrt{\Delta}$ once at $\lambda_I$. 
\end{lemma}
\begin{proof}
It is not difficult to see that $B$ does not intersect the right branch of $\sqrt{\Delta}$ for both (i) and (ii). $\sqrt{\Delta}$ is linear in $\lambda$ with slope
$a^2-1$ for large $\lambda$. The slope of $B$ is $a^2+1$. Both
$a^2-1$ and $a^2+1$ are positive and $a^2+1>a^2-1$, thus $B$
and the right branch of $\sqrt{\Delta}$ never meet. When $\lambda_l<\lambda_B<\lambda_r$, $B<0$ for $\lambda<\lambda_B$ and $\sqrt{\Delta}>0$ for $\lambda<\lambda_l<\lambda_B$. Therefore $B$ and $\sqrt{\Delta}$ never intersect. 
In (ii), $B$ intersects the left branch of $\sqrt{\Delta}$ at
$\lambda_I={\displaystyle \frac{2A\alpha-2a \alpha-a}{a}}$.  
\end{proof}

\section*{Acknowledgments}
We thank G.~Bard~Ermentrout, William~Troy, Xinfu~Chen, Jonathan~Rubin,
and Bjorn~Sandestade for illuminating discussions.  This work was
supported by the National Institute of Mental Health and the
A.~P.~Sloan foundation.

\nocite{Aliprantis0,Aliprantis1,Arbib,Atkinson,Bender,Boyce,Champneys1,Delevs,Duffy,Ellias,Enander,Ermentrout1,Ermentrout2,Folland,Fuster1,Garvan,Green,Griffel,Gutkin,Haskell,Kato,Kuznetsov,Laing1,Laing2,Laing3,Miller,Morrison,Murray,Nicholls,Nishiura,Pelinovsky,Polianin,Powers,Rahman,Rubin,Rubin1,Rudin,Salinas,Strogatz,Troy2,Troybook,Wiggins,Wilson1,XPPAUT,Zeidler}
 
\bibliographystyle{plain}
\bibliography{paper2}

\begin{thebibliography}{10}

\bibitem{Aliprantis1}
C.~D. Aliprantis.
\newblock {\em Problems in real analysis: a workbook with solutions}.
\newblock Academic Press, 1999.

\bibitem{Aliprantis0}
C.~D. Aliprantis and Burkinshaw.
\newblock {\em Principles of real analysis}.
\newblock Academic Press, 1998.

\bibitem{amari77}
S.~Amari.
\newblock Dynamics of pattern formation in lateral-inhibition type neural
  fields.
\newblock {\em Biol. Cybernetics}, 27:77--87, 1977.

\bibitem{Arbib}
M.~A. Arbib, editor.
\newblock {\em The Handbook of Brain Theory and Neural Networks}.
\newblock MIT Press, 1995.

\bibitem{Atkinson}
K.~E. Atkinson.
\newblock {\em Numerical solution of integral equations of the second kind}.
\newblock Cambridge University Press, 1997.

\bibitem{Bender}
C.~M. Bender and S.~A. Orszag.
\newblock {\em Advanced mathematical methods for scientists and engineers I:
  Asyptotic methods and perturbation theory}.
\newblock Springer, 1999.

\bibitem{Boyce}
W.~E. Boyce and R.~C. DiPrima.
\newblock {\em Introduction to differential equations}.
\newblock John Wiley and Sons, 1970.

\bibitem{Champneys1}
A.~R. Champneys and J.~P. McKenna.
\newblock On solitary waves of a piece-wise linear suspended beam model.
\newblock {\em Nonlinearity}, 10:1763--1782, 1997.

\bibitem{Delevs}
L.~M. Delves and J.~L. Mohamed.
\newblock {\em Computational methods for integral equations}.
\newblock Cambridge University Press, 1988.

\bibitem{Duffy}
D.~G. Duffy.
\newblock {\em Green's functions with applications}.
\newblock Chapman and Hall/CRC, 2001.

\bibitem{Ellias}
S.~A. Ellias and S.~Grossberg.
\newblock Pattern formation, contrast control, and oscillations in the
  short-term memory of shunting on-center off-surround networks.
\newblock {\em Biol. Cybern.}, 20:69--98, 1975.

\bibitem{XPPAUT}
G.~B. Ermentrout.
\newblock Xppaut, simulation software tool.

\bibitem{Ermentrout1}
G.~B. Ermentrout.
\newblock Neural networks as spatio-temporal pattern-forming systems.
\newblock {\em Rep. Prog. Phys.}, 61:353--430, 1998.

\bibitem{Ermentrout2}
G.~B Ermentrout.
\newblock {\em Simulating, Analyzing, and Animating Dynamical Systems: A Guide
  to XPPAUT for Researchers and Students}.
\newblock SIAM, 2002.

\bibitem{Evans}
J.~W. Evans.
\newblock Nerve axon equations, i: Linear approximations.
\newblock {\em Indiana Univ. Math. J.}, 21:877--955, 1972.

\bibitem{E1}
J.~W. Evans.
\newblock Nerve axon equations, ii: Stability at rest.
\newblock {\em Indiana Univ. Math. J.}, 22:75--90, 1972.

\bibitem{E2}
J.~W. Evans.
\newblock Nerve axon equations, iii: Stability of the nerve impulse.
\newblock {\em Indiana Univ. Math. J.}, 22:577--594, 1972.

\bibitem{E3}
J.~W. Evans.
\newblock Nerve axon equations, iv: The stable and unstable impulse.
\newblock {\em Indiana Univ. Math. J.}, 24:1169--1190, 1975.

\bibitem{Folland}
G.~B. Folland.
\newblock {\em Fourier analysis and its applications}.
\newblock Wadsworth and Brooks/Cole Advanced Books and Software, 1992.

\bibitem{Fuster1}
J.~M Fuster.
\newblock {\em Prefrontal cortex: anatomy, physiology, and neuropsychology of
  the frontal lobe}.
\newblock Lippincott-Raven Publishers, 1997.

\bibitem{Garvan}
F.~Garvan.
\newblock {\em The Maple Book}.
\newblock Chapman and Hall, 2001.

\bibitem{Green}
C.~D. Green.
\newblock {\em Integral equation methods}.
\newblock Nelson, 1969.

\bibitem{Griffel}
D.~H. Griffel.
\newblock {\em Applied functional analysis}.
\newblock Ellis Horwood, 1985.

\bibitem{Guo1}
Y.~Guo.
\newblock Existence and stability of standing pulses in neural networks.
\newblock PhD thesis, University of Pittsburgh, 2003.

\bibitem{Guo2}
Y.~Guo and C.C. Chow.
\newblock Existence and stability of standing pulses in neural networks: I.
  existence.
\newblock Preprint, 2004.

\bibitem{Gutkin}
B.S. Gutkin, C.R. Laing, C.~L. Colby, C.C. Chow, and G.~B. Ermentrout.
\newblock Turing on and off with excitation: the role of spike-timing
  asynchrony and synchrony in sustained neural activity.
\newblock {\em J. Comp. Neurosci.}, 11:121--134, 2001.

\bibitem{Haskell}
E~Haskell and P.~C. Bressloff.
\newblock On the formation of persistent states in neuronal networks models of
  feature selectivity.
\newblock {\em J. Integ. Neurosci.}, 2:103--123, 2003.

\bibitem{Kato}
T.~Kato.
\newblock {\em Perturbation theory for linear operators}.
\newblock Springer, 1995.

\bibitem{Functional3}
E.~Kreyszig.
\newblock {\em Introductory Functional Analysis with Applications}.
\newblock John Wiley and Sons, 1978.

\bibitem{Kuznetsov}
Y.~A. Kuznetsov.
\newblock {\em Elements of applied bifurcation theory}.
\newblock Springer, 1998.

\bibitem{Laing2}
C.~R. Laing and C.~C. Chow.
\newblock Stationary bumps in networks of spiking neurons.
\newblock {\em Neural Comp.}, 13:1473--1493, 2001.

\bibitem{Troy2}
C.~R. Laing and W.~C. Troy.
\newblock Two-bump solutions of amari type models of working memory.
\newblock {\em Physica D}, pages 190--218, 2003.

\bibitem{Laing1}
C.~R. Laing, w.~C. Troy, B~Gutkin, and G.~B Ermentrout.
\newblock Multiple bumps in a neuronal model of working memory.
\newblock {\em SIAM J. of Applied Math.}, 63, no.1:62--97, 2002.

\bibitem{Laing3}
C.~R. Laing and William~C. Troy.
\newblock Pde methods for nonlocal models.
\newblock {\em SIAM Journal on Applied Dynamical Systems}, 2:487--516, 2001.

\bibitem{Miller}
E.~K. Miller, C.~A. Erickson, and R.~Desimone.
\newblock Neural mechanisms of visual working memory in prefrontal cortex of
  the macaque.
\newblock {\em J. Neurosci.}, 16:5154--5167, 1996.

\bibitem{Morrison}
N.~Morrison.
\newblock {\em Introduction to Fourier analysis}.
\newblock Wiley-Interscience, 1994.

\bibitem{Murray}
J.~D. Murray.
\newblock {\em Mathematical biology}.
\newblock Springer, 2002.

\bibitem{Nicholls}
J.~G. Nicholls.
\newblock {\em From neuron to brain: a cellular molecular approach to the
  function of the nervous system}.
\newblock Sinauer Associates, 1992.

\bibitem{Nishiura}
Y.~Nishiura and M.~Mimura.
\newblock Layer oscillations in reactoin-diffusion systems.
\newblock {\em SIAM J. Appl. Math.}, 49:481--514, 1989.

\bibitem{Enander}
E.~P$\ddot{a}$rt-Enander, A.~Sj$\ddot{o}$berg, Melin B., and Isaksson P.
\newblock {\em The MATLAB handbook}.
\newblock Addison-Wesley, 1998.

\bibitem{Troybook}
L.~A. Peletier and W.~C. Troy.
\newblock {\em Spatial patterns: higher order models in physics and mechanics}.
\newblock Birkhauser, 2001.

\bibitem{Pelinovsky}
E.~Pelinovsky, D and V.~G. Yakhno.
\newblock Generation of collective-activity structures in a homogeneous
  neuron-like medium. i. bifurcation analysis of static structures,.
\newblock {\em Bifurcation Chaos Appl. Sci. Eng.}, 6:81--87, 89--100, 1996a,b.

\bibitem{Pinto1}
J.~D. Pinto and Ermentrout~G. B.
\newblock Spatially structured activity in synaptically coupled neuronal
  networks:2 lateral inhibition and standing pulses.
\newblock {\em SIAM J. Appl. Math.}, 62:226--243, 2001.

\bibitem{Polianin}
A.~D. Polianin and A.~V. Manzhirov.
\newblock {\em Handbook of integral equations}.
\newblock CRC Press, 1998.

\bibitem{Powers}
D.~L. Powers.
\newblock {\em Boundary value problems}.
\newblock Harcourt Academic Press, 1999.

\bibitem{Rahman}
M.~Rahman.
\newblock {\em Complex variables and transform calculus}.
\newblock Computational Mechanics Publications, 1997.

\bibitem{Rubin}
J.~E. Rubin, D.~Terman, and C.~C. Chow.
\newblock Localized bumps of activity sustained by inhibition in a two-layer
  thalamic network.
\newblock {\em J Comp Neurosci}, 10:313--331, 2001.

\bibitem{Rubin1}
J.~E Rubin and W.~C Troy.
\newblock Sustained spatial patterns of activity in neuronal populations with
  or without lateral inhibition.
\newblock {\em SIAM J. Appl. Math.}, 2004.

\bibitem{Rudin}
W.~Rudin.
\newblock {\em Principles of mathematical analysis}.
\newblock McGraw-Hill, 1976.

\bibitem{Salinas}
E.~Salinas and L.~F. Abbott.
\newblock A model of multiplicative neural responses in parietal cortex.
\newblock {\em Proc Natl Acad Sci USA}, 93:11956--11961, 1996.

\bibitem{seung}
S.~H. Seung.
\newblock How the brain keeps the eyes still.
\newblock {\em Proc Natl Acad Sci USA}, 93:13339--44, 1996.

\bibitem{Strogatz}
S.~H. Strogatz.
\newblock {\em Nonlinear dynamics and chaos}.
\newblock Perseus Books, 1994.

\bibitem{Wiggins}
S.~Wiggins.
\newblock {\em Introduction to applied nonlinear dynamical systems and chaos}.
\newblock Springer, 1990.

\bibitem{Wilson1}
H.~R. Wilson and J.~D. Cowan.
\newblock A mathematical theory of the functional dynamics of cortical and
  thalamic nervous tissue.
\newblock {\em Kybernetic}, 13:55--80, 1973.

\bibitem{Mathematica}
S.~Wolfram.
\newblock {\em The Mathematica Book}.
\newblock Cambridge University Press, 4th Edition, 1999.

\bibitem{Zeidler}
E.~Zeidler.
\newblock {\em Nonlinear functional analysis and its applications I:
  fixed-point theorems}.
\newblock Springer, 1986.

\end{thebibliography}

\end{document}